\documentclass[11pt, leqno]{article}

\usepackage{amsmath, amsthm, amssymb, setspace, fullpage, graphicx, cite}
\usepackage{color}

\title{\huge Max-Weight Scheduling in Queueing Networks with Heavy-Tailed Traffic
\footnote{This work was supported by NSF Grants CNS-0915988 and CCF-0728554, and ARO MURI Grant W911NF-08-1-0238.}%
}

\author{Mihalis G. Markakis, Eytan H. Modiano, and John N. Tsitsiklis
\thanks{The authors are with the Laboratory for Information and Decision Systems, at the Massachusetts Institute of Technology, Cambridge, MA, USA.}%
}

\begin{document}

\date{}

\maketitle

\begin{abstract}
We consider the problem of packet scheduling in single-hop queueing networks, and analyze the impact of heavy-tailed traffic on the performance of Max-Weight scheduling. As a performance metric we use the delay stability of traffic flows: a traffic flow is delay stable if its expected steady-state delay is finite, and delay unstable otherwise. First, we show that a heavy-tailed traffic flow is delay unstable under any scheduling policy. Then, we focus on the celebrated Max-Weight scheduling policy, and show that a light-tailed flow that conflicts with a heavy-tailed flow is also delay unstable. This is true irrespective of the rate or the tail distribution of the light-tailed flow, or other scheduling constraints in the network. Surprisingly, we show that a light-tailed flow can be delay unstable, even when it does not conflict with heavy-tailed traffic. Furthermore, delay stability in this case may depend on the rate of the light-tailed flow. Finally, we turn our attention to the class of Max-Weight-$\alpha$ scheduling policies; we show that if the $\alpha$-parameters are chosen suitably, then the sum of the $\alpha$-moments of the steady-state queue lengths is finite. We provide an explicit upper bound for the latter quantity, from which we derive results related to the delay stability of traffic flows, and the scaling of moments of steady-state queue lengths with traffic intensity.
\end{abstract}


\section{Introduction}

\par We study the impact of heavy-tailed traffic on the performance of scheduling policies in single-hop queueing networks. Single-hop network models have been used extensively to capture the dynamics and scheduling decisions in real-world communication networks, such as wireless uplinks and downlinks, switches, wireless ad hoc networks, sensor networks, and call centers. In all these systems, one cannot serve all queues simultaneously, e.g., due to wireless interference constraints, giving rise to a scheduling problem. Clearly, the overall performance of the network depends critically on the scheduling policy applied.

\par The focus of this paper is on a well-studied class of scheduling policies, commonly refered to as Max-Weight policies. This class of policies was introduced in the seminal work of Tassiulas and Ephremides \cite{TE92}, and since then numerous studies have analyzed the performance of such policies in different settings, e.g., see \cite{AKRSVW04,GNT06}, and the references therein. A remarkable property of Max-Weight policies is their \textbf{throughput optimality}, i.e., their ability to stabilize a queueing network whenever this is possible, without any information on the arriving traffic. Moreover, it has been shown that policies from this class achieve low, or even optimal, average delay for specific network topologies, when the arriving traffic is light-tailed \cite{GMT07,N08,S04,SW06,TE93}. \footnote{On the other hand, when Max-Weight scheduling is combined with Back-Pressure routing in the context of multi-hop networks, there is evidence that delay performance can be poor, e.g., see the discussion in \cite{BSS09}.} However, the performance of Max-Weight scheduling in the presence of heavy-tailed traffic is not well understood.

\par We are motivated to study networks with heavy-tailed traffic by significant evidence that traffic in real-world communication networks exhibits strong correlations and statistical similarity over different time scales. This observation was first made by Leland \emph{et al.} \cite{LTWW94} through analysis of Ethernet traffic traces. Subsequent empirical studies have documented this phenomenon in other networks, while accompanying theoretical studies have associated it with arrival processes that have heavy tails; see \cite{PW00} for an overview. The impact of heavy tails has been analyzed extensively in the context of single or multi-server queues; see the survey papers \cite{BBNZ03,BZ07}, and the references therein. However, the related work is rather limited in the context of queueing networks, e.g., see the paper by Borst \emph{et al.} \cite{BMU03}, which studies the ``Generalized Processor Sharing'' policy.

\par This paper aims to fill a gap in the literature, by analyzing the impact of heavy-tailed traffic on the performance of Max-Weight scheduling in single-hop queueing networks. In particular, we study the delay stability of traffic flows: a traffic flow is delay stable if its expected steady-state delay is finite, and delay unstable otherwise. Our previous work \cite{MMT09} gives some preliminary results in this direction, in a simple system with two parallel queues and a single server. The \textbf{main contributions} of this paper include: i) in a single-hop queueing network under the Max-Weight scheduling policy, we show that any light-tailed flow that conflicts with a heavy-tailed flow is delay unstable; ii) surprisingly, we also show that for certain admissible arrival rates, a light-tailed flow can be delay unstable even if it does not conflict with heavy-tailed traffic; iii) we analyze the Max-Weight-$\alpha$ scheduling policy, and show that if the $\alpha$-parameters are chosen suitably, then the sum of the $\alpha$-moments of the steady-state queue lengths is finite. We use this result to prove that by proper choice of the $\alpha$-parameters, all light-tailed flows are delay stable. Moreover, we show that Max-Weight-$\alpha$ achieves the optimal scaling of higher moments of steady-state queue lengths with traffic intensity.

\par The rest of the paper is organized as follows. Section 2 contains a detailed presentation of the model that we analyze, namely, a single-hop queueing network. It also defines formally the notions of heavy-tailed and light-tailed traffic, and of delay stability. In Section 3 we motivate the subsequent development by presenting, informally and through simple examples, the main results of the paper. In Section 4 we analyze the performance of the celebrated Max-Weight scheduling policy. Our general results are accompanied by examples, which illustrate their implications in practical network settings. Section 5 contains the analysis of the parameterized Max-Weight-$\alpha$ scheduling policy, and the performance that it achieves in terms of delay stability. This section also includes results about the scaling of moments of steady-state queue lengths with the traffic intensity and the size of the network, accompanied by several examples. We conclude with a discussion of our findings and future research directions in Section 6. The appendices contain some background material and most of the proofs of our results.


\bigskip\section{Model and Problem Formulation}

\par We start with a detailed presentation of the queueing model considered in this paper, together with some necessary definitions and notation. 

\par We denote by $\Re_+$, $Z_+$, and $N$ the sets of nonnegative reals, nonnegative integers, and positive integers, respectively. The cartesian products of $M$ copies of $\Re_+$ and $Z_+$ are denoted by $\Re_+^M$ and $Z_+^M$, respectively.

\par We assume that time is slotted and that arrivals occur at the end of each time slot. The topology of the network is captured by a directed graph $G=(\cal{N},\cal{E})$, where $\cal{N}$ is the set of nodes and $\cal{E}$ is the set of (directed) edges. Our model involves single-hop traffic flows: data arrives at the source node of an edge, for transmission to the node at the other end of the edge, where it exits the network. More formally, let $F \in N$ be the number of traffic flows of the network. A \textbf{traffic flow} $f \in \{1,\ldots,F\}$ consists of a discrete time stochastic arrival process $\{A_f(t);\ t \in Z_+\}$, a source node $s(f)$, and a destination node $d(f)$, with $s(f),d(f) \in \cal{N}$, and $(s(f),d(f)) \in \cal{E}$. We assume that each arrival process $\{A_f(t);\ t \in Z_+\}$ takes values in $Z_+$, and is independent and identically distributed (IID) over time. Furthermore, the arrival processes associated with different traffic flows are mutually independent. We denote by $\lambda_f=E[A_f(0)]>0$ the rate of traffic flow $f$, and by $\lambda = (\lambda_f;\ f=1,\ldots,F)$ the vector of the rates of all traffic flows.

\medskip\par\textbf{Definition 1: (Heavy Tails)} A traffic flow $f$ is heavy-tailed if $E[A_f^2(0)]=\infty$, and light-tailed otherwise.

\medskip\par The traffic of flow $f$ is buffered in a dedicated queue at node $s(f)$ (queue $f$, henceforth.) Our modeling assumptions imply that the set of traffic flows can be identified with the set of edges and the set of queues of the network. The service discipline within each queue is assumed to be ``First Come, First Served.'' The stochastic process $\{Q_f(t);\ t \in Z_+\}$ captures the evolution of the length of queue $f$. Since our motivation comes from communication networks, $A_f(t)$ will be interpreted as the number of packets that queue $f$ receives at the end of time slot $t$, and $Q_f(t)$ as the total number of packets in queue $f$ at the beginning of time slot $t$. The arrivals and the lengths of the various queues at time slot $t$ are captured by the vectors $A(t)=(A_f(t);\ f=1,\ldots,F)$ and $Q(t)=(Q_f(t);\ f=1,\ldots,F)$, respectively.

\par In the context of a communication network, a batch of packets arriving to a queue at any given time slot can be viewed as a single entity, e.g., as a file that needs to be transmitted. We define the \textbf{end-to-end delay of a file} of flow $f$ to be the number of time slots that the file spends in the network, starting from the time slot right after it arrives at $s(f)$, until the time slot that its last packet reaches $d(f)$. For $k \in N$, we denote by $D_f(k)$ the end-to-end delay of the $k^{th}$ file of queue $f$. The vector $D(k)=(D_f(k);\ f=1,\ldots,F)$ captures the end-to-end delay of the $k^{th}$ files of the different traffic flows.

\par In general, not all edges can be activated simultaneously, e.g., due to interference in wireless networks, or matching constraints in a switch. Consequently, not all traffic flows can be served simultaneously. A set of traffic flows that can be served simultaneously is called a \textbf{feasible schedule}. We denote by $S$ the set of all feasible schedules, which is assumed to be an arbitrary subset of the powerset of $\{1,\ldots,F\}$. For simplicity, we assume that all attempted transmissions of data are successful, that all packets have the same size, and that the transmission rate along any edge is equal to one packet per time slot. We denote by $S_f(t) \in \{0,1\}$ the number of packets that are scheduled for transmission from queue $f$ at time slot $t$. Note that this is not necessarily equal to the number of packets that are transmitted because the queue may be empty.

\par Let us now define formally the notion of a \textbf{scheduling policy}. The past history and present state of the system at time slot $t \in N$ is captured by the vector
\begin{equation}
H(t) = (Q(0),A(0),\ldots,Q(t-1),A(t-1),Q(t)). \nonumber
\end{equation}
\noindent At time slot 0, we have $H(0)=(Q(0))$. A (causal) scheduling policy is a sequence $\pi=(\mu_0,\mu_1,\ldots)$ of functions $\mu_t: H(t) \to S,\ t \in Z_+$, used to determine scheduling decisions, according to $S(t)=\mu_t(H(t))$.

\par Using the notation above, the \textbf{dynamics} of queue $f$ take the form:
\begin{equation}
Q_f(t+1) = Q_f(t) + A_f(t) - S_f(t) \cdot 1_{\{Q_f(t)>0\}}, \nonumber
\end{equation}
\noindent for all $t \in Z_+$, where $1_{\{Q_f(t)>0\}}$ denotes the indicator function of the event $\{Q_f(t)>0\}$. The vector of initial queue lengths $Q(0)$ is assumed to be an arbitrary element of $Z_+^F$.

\par We restrict our attention to scheduling policies that are \textbf{regenerative}, i.e., policies under which the network starts afresh probabilistically in certain time slots. More precisely, under a regenerative policy there exists a sequence of stopping times $\{\tau_n;\ n \in Z_+\}$
with the folowing properties. i) The sequence $\{\tau_{n+1}-\tau_n;\ n \in Z_+\}$ is IID. ii) Let $X(t)=(Q(t),A(t),S(t))$, and consider the processes that describe the ``cycles'' of the network, namely, $C_0=\{X(t);\ 0 \leq t<\tau_0\}$, and $C_n=\{X(\tau_{n-1}+t);\ 0 \leq t<\tau_n-\tau_{n-1}\},\ n \in N$; then, $\{C_n;\ n \in N\}$ is an IID sequence, independent of $C_0$. iii) The (lattice) distribution of the cycle lengths, $\tau_{n+1}-\tau_n$, has span equal to one and finite expectation.

\par Properties (i) and (ii) imply that the queueing network evolves like a (possibly delayed) regenerative process. Property (iii) states that this process is aperiodic and positive recurrent, which will be crucial for the stability of the network. The following definition gives the precise notion of stability that we use in this paper.

\medskip\par\textbf{Definition 2: (Stability)} The single-hop queueing network described above is stable under a specific scheduling policy, if the vector-valued sequences $\{Q(t);\ t \in Z_+\}$ and $\{D(k);\ k \in N\}$ converge in distribution, and their limiting distributions do not depend on the initial queue lengths $Q(0)$.

\medskip\par Notice that our definition of stability is slightly different than the commonly used definition (positive recurrence of the Markov chain of queue lengths), since it includes the convergence of the sequence of file delays $\{D(k);\ k \in N\}$. The reason is that in this paper we study properties of the limiting distribution of $\{D(k);\ k \in N\}$ and, naturally, we need to ensure that this limiting distribution exists.

\par Under a stabilizing scheduling policy, we denote by $Q=(Q_f;\ f=1,\ldots,F)$ and $D=(D_f;\ f=1,\ldots,F)$ the limiting distributions of $\{Q(t);\ t \in Z_+\}$ and $\{D(k);\ k \in N\}$, respectively. The dependence of these limiting distributions on the scheduling policy has been suppressed from the notation, but will be clear from the context. We refer to $Q_f$ as the steady-state length of queue $f$. Similarly, we refer to $D_f$ as the steady-state delay of a file of traffic flow $f$. We note that under a regenerative policy (if one exists), the queueing network is guaranteed to be stable. This is because the sequences of queue lengths and file delays are (possibly delayed) aperiodic and positive recurrent regenerative processes, and, hence, converge in distribution; see \cite{SW93}.

\par The stability of the queueing network depends on the rates of the various traffic flows relative to the transmission rates of the edges and the scheduling constraints. This relation is captured by the stability region of the network.

\medskip\par\textbf{Definition 3: (Stability Region)} \cite{TE92} The stability region of the single-hop queueing network described above, denoted by $\Lambda$, is the set of rate vectors:
\begin{equation}
\Big\{ \lambda \in \Re_+^F \ \Big| \ \exists \ \zeta_s \in \Re_+,\ s \in S: \ \lambda \leq \sum_{s \in S} \zeta_s \cdot s,\ \sum_{s \in S} \zeta_s < 1 \Big\}. \nonumber
\end{equation}

\medskip\par In other words, a rate vector $\lambda$ belongs to $\Lambda$ if there exists a convex combination of feasible schedules that covers the rates of all traffic flows. If a rate vector is in the stability region of the network, then the traffic corresponding to this vector is called \textbf{admissible}, and there exists a scheduling policy under which the network is stable.

\medskip\par\textbf{Definition 4: (Traffic Intensity)} The traffic intensity of a rate vector $\lambda \in \Lambda$ is a real number in [0,1) defined as:
\begin{equation}
\rho(\lambda) = \inf \Big\{ \sum_{s \in S} \zeta_s \ \Big| \ \lambda \leq \sum_{s \in S} \zeta_s \cdot s,\ \zeta_s \in \Re_+,\ \forall s \in S \Big\}. \nonumber
\end{equation}

\medskip\par Clearly, arriving traffic with rate vector $\lambda$ is admissible if and only if $\rho(\lambda)<1$. \textbf{Throughout this paper we assume that the traffic is admissible}. 

\par Let us now define the property that we use to evaluate the performance of scheduling policies, namely, the delay stability of a traffic flow. 

\medskip\par\textbf{Definition 5: (Delay Stability)} A traffic flow $f$ is delay stable under a specific scheduling policy if the queueing network is stable under that policy and $E[D_f]<\infty$; otherwise, the traffic flow $f$ is delay unstable.

\medskip\par The following lemma relates the steady-state quantities $E[Q_f]$ and $E[D_f]$, and will help us prove delay stability results.

\medskip\par\textbf{Lemma 1:} Consider the single-hop queueing network described above under a regenerative scheduling policy. Then,
\begin{equation}
E[Q_f]<\infty \ \Longleftrightarrow \ E[D_f]<\infty, \qquad \forall f \in \{1,\ldots,F\}. \nonumber
\end{equation}

\begin{proof}
see Appendix 1.1.
\end{proof}

\medskip\par\textbf{Theorem 1: (Delay Instability of Heavy Tails)} Consider the single-hop queueing network described above under a regenerative scheduling policy. Every heavy-tailed traffic flow is delay unstable.

\begin{proof}
(Sketch) The result follows easily from the Pollaczek-Khinchine formula for the expected delay in a $M/G/1$ queue, and a stochastic comparison argument. The main idea is that in a heavy-tailed traffic flow, the probability that a very big file arrives to the respective queue is relatively high. Combined with the ``First Come, First Served'' discipline within the queue, this implies that a large number of files, arriving after the big one, experience very large delays. This is true even if the queue gets served whenever it is nonempty, namely, if the queue is given preemptive priority. Consequently, under any scheduling policy, there is relatively high probability that a large number of files experiences very large delays. This then implies that a heavy-tailed traffic flow is delay unstable. For a formal proof see Appendix 2.
\end{proof}

\medskip\par Since there is little we can do about the delay stability of heavy-tailed flows, we turn our attention to light-tailed traffic. The Pollaczek-Khinchine formula for the expected delay in a $M/G/1$ queue implies that the intrinsic burstiness of light-tailed traffic is not sufficient to cause delay instability. However, scheduling in a queueing network couples the statistics of different traffic flows. We will see that this coupling can cause light-tailed flows to become delay unstable, giving rise to a form of \textbf{propagation of delay instability}.


\bigskip\section{Overview of Main Results}

\par In this section we introduce, informally and through simple examples, the main results of the paper and the basic intuition behind them.

\par Let us start with the queueing system of Figure 1, which consists of two parallel queues and a single server. Traffic flow 1 is assumed to be heavy-tailed, whereas traffic flow 2 is light-tailed. Service is allocated according to the Max-Weight scheduling policy, which is equivalent to ``Serve the Longest Queue'' in this simple setting. Theorem 1 implies that traffic flow 1 is delay unstable. Our findings imply that \textbf{traffic flow 2 is also delay unstable, even though it is light-tailed}. The intuition behind this result is that queue 1 is occasionally very long (infinite, in steady-state expectation) because of its heavy-tailed arrivals. When this happens, and under the Max-Weight policy, queue 2 has to build up to a similar length in order to receive service. A very long queue then implies very large delays for the files of that queue under ``First Come, First Served,'' which leads to delay instability.
 
\begin{figure}[ht]
\centering
\includegraphics[scale=0.45]{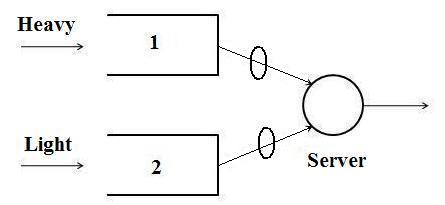}
\caption{Delay instability in parallel queues with heavy-tailed traffic.}\label{fig:2queues}
\end{figure}

\par Systems of parallel queues have been analyzed extensively in the literature. One of the main reasons is that their simple dynamics often lead to elegant analysis and clean results. However, real-world communication networks are much more complex. In this paper we go beyond parallel queues and analyze queueing networks with more complicated structure. A simple example is the queueing network of Figure 2, where traffic flow 1 is assumed to be heavy-tailed, whereas traffic flows 2 and 3 are light-tailed. The server can serve either queue 1 alone, or queues 2 and 3 simultaneously. This example could represent a wireless network with interference constraints. In this setting the Max-Weight policy compares the length of queue 1 to the sum of the lengths of queues 2 and 3, and serves the ``heavier'' schedule. 

\begin{figure}[ht]
\centering
\includegraphics[scale=0.45]{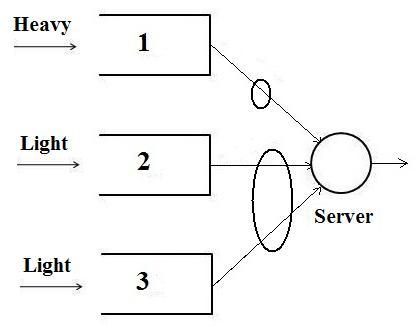}
\caption{Propagation of delay instability: conflicting with heavy-tailed traffic.}\label{fig:3queues1}
\end{figure}

\par The intuition from the previous example suggests that at least one of the queues 2 and 3 has to build up to the order of magnitude of queue 1, in order for these two queues to receive service. In other words, we expect that at least one of the traffic flows 2 and 3 will be delay unstable under Max-Weight. Our findings imply that, in fact, \textbf{both traffic flows are delay unstable}. The main idea behind this result is the following: with positive probability, the arrival processes to queues 2 and 3 exhibit their ``average'' behavior. In that case, the corresponding queues build up slowly and together, which implies that when they claim the server they have both built up to the order of magnitude of queue 1. 

\par The simple networks of Figures 1 and 2 illustrate special cases of a general result: every light-tailed flow that conflicts with a heavy-tailed flow is delay unstable. For more details see Theorem 2 in Section 4.1.

\begin{figure}[ht]
\centering
\includegraphics[scale=0.45]{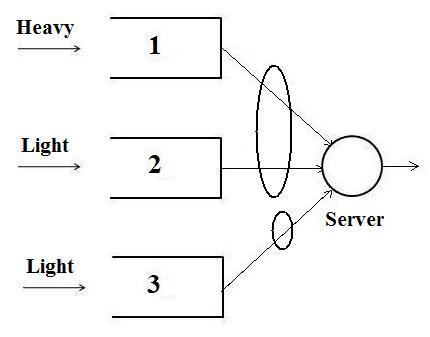}
\caption{Propagation of delay instability: concurring with heavy-tailed traffic.}\label{fig:3queues2}
\end{figure}

\par Going one step further, consider the queueing network of Figure 3. Traffic flow 1 is assumed to be heavy-tailed, whereas traffic flows 2 and 3 are light-tailed. The server can serve either queues 1 and 2 simultaneously, or queue 3 alone. In this setting the Max-Weight policy compares the length of queue 3 to the sum of the lengths of queues 1 and 2, and serves the ``heavier'' schedule. The intuition from the previous examples suggests that traffic flow 3 is delay unstable, but the real question is the delay stability of traffic flow 2. One would expect that this flow is delay stable: it is light-tailed itself, and is served together with a heavy-tailed flow, which should result in more service opportunities under Max-Weight. Surprisingly though, we show that \textbf{there exist arrival rates within the stability region of this network, such that traffic flow 2 is delay unstable}. The key observation here is that even though traffic flow 2 does not conflict with heavy-tailed traffic, it does conflict with traffic flow 3, which is delay unstable because it conflicts with heavy-tailed traffic. For more details see Propositions 1, 3, and 4 in Sections 4.2 and 4.3.

\par The examples above suggest that in queueing networks with heavy-tailed traffic, delay instability not only appears but propagates through the network under the Max-Weight policy. Seeking a remedy to this situation, we turn to the more general Max-Weight-$\alpha$ scheduling policy. This policy assigns a positive $\alpha$-parameter to each traffic flow, and instead of comparing the lengths of the queues/schedules, and serving the longest one, it compares the lengths of the queues to the respective $\alpha$-powers. Our findings imply that in the network of Figure 1, we can guarantee that \textbf{traffic flow 2 is delay stable, provided the $\alpha$-parameter for traffic flow 1 is sufficiently small}. In other words, we prevent the propagation of delay instability. This is a special case of a general result: if the $\alpha$-parameters of the Max-Weight-$\alpha$ policy are chosen suitably, then the sum of the $\alpha$-moments of the steady-state queue lengths is finite. For more details see Theorem 3 in Section 5.1.


\bigskip\section{Max-Weight Scheduling}

\par In this section we evaluate the performance of the Max-Weight scheduling policy, with respect to the delay stability of traffic flows. Informally speaking, the ``weight'' of a feasible schedule is the sum of the lengths of all queues included in it. As its name suggests, the Max-Weight policy activates a feasible schedule with the maximum weight at any given time slot. More formally, under the Max-Weight policy, the scheduling vector $S(t)$ belongs to the set:
\begin{equation}
S(t) \ \in \ \arg\max_{(s_f) \in S} \Big\{ \sum_{f=1}^F Q_f(t) \cdot s_f \Big\}. \nonumber
\end{equation}

\par If this set includes multiple feasible schedules, then one of them is chosen uniformly at random. The following lemma states that the network is stable under the Max-Weight policy. Essentially, this result is well-known, e.g., for light-tailed traffic, see \cite{TE92}; for more general arrivals, see \cite{S04}. A subtle point is that in this paper we adopt a somewhat different definition for stability. So, we have to ensure that, apart from the sequences of queue lengths, the sequences of file delays converge as well.

\medskip\par\textbf{Lemma 2: (Stability under Max-Weight)} The single-hop queueing network described in Section 2 is stable under the Max-Weight scheduling policy. 

\begin{proof}
Consider the single-hop queueing network of Section 2 under the Max-Weight scheduling policy. It can be verified that the sequence $\{Q(t);\ t \in Z_+\}$ is a time-homogeneous, irreducible, and aperiodic Markov chain on the countable state-space $Z_+^F$. Proposition 2 of \cite{S04} implies that this Markov chain is also positive recurrent. Hence, $\{Q(t);\ t \in Z_+\}$ converges in distribution, and its limiting distribution does not depend on $Q(0)$. Based on this, it can be verified that the sequence $\{D(k);\ k \in N\}$ is a (possibly delayed) aperiodic and positive recurrent regenerative process. Therefore, it also converges in distribution, and its limiting distribution does not depend on $Q(0)$; see \cite{SW93}.
\end{proof}


\bigskip\subsection{Conflicting with Heavy-Tailed Flows}

\par In this section we state one of the main results of the paper, which generalizes our observations from the simple networks of Figures 1 and 2. Before we give the result, though, let us define precisely the notion of conflict between traffic flows.

\medskip\par\textbf{Definition 6:} The traffic flow $f$ conflicts with $f'$, and vice versa, if there exists no feasible schedule in $S$ that includes both $f$ and $f'$. 

\medskip\par\textbf{Theorem 2: (Conflicting with Heavy Tails)} Consider the single-hop queueing network described in Section 2 under the Max-Weight scheduling policy. Every light-tailed flow that conflicts with a heavy-tailed flow is delay unstable.

\begin{proof}
(Sketch) Let $h$ and $l$ be a heavy-tailed and a light-tailed traffic flow, respectively, and suppose that $l$ conflicts with $h$. Queue $h$ is occasionally very long (infinite, in steady-state expectation), due to the heavy-tailed nature of the traffic that it receives. In order for queue $l$ to get served, the weight of at least one feasible schedule that includes $l$ has to build up to the order of magnitude of queue $h$. However, with positive probability, the arrival processes of all feasible schedules that include $l$ exhibit their ``average'' behavior. In that case, queue $l$ builds up at a roughly constant rate, for a time period of the order of magnitude of queue 1. Combined with Lemma 1, this implies that traffic flow $l$ is delay unstable. For a formal proof see Appendix 3.
\end{proof}

\medskip\par We emphasize the generality of this result. Namely, a light-tailed flow that conflicts with heavy-tailed traffic is delay unstable, irrespective of: i) its rate; ii) the tail asymptotics of its underlying distribution; iii) whether it is scheduled alone or with other traffic flows. Hence, we view Theorem 2 as capturing a ``universal phenomenon'' for the propagation of delay instability.


\medskip\subsection{Concurring with Heavy-Tailed Flows}

\par So far we have shown that: i) a heavy-tailed traffic flow is delay unstable under any regenerative scheduling policy; and ii) a light-tailed traffic flow that conflicts with a heavy-tailed flow is delay unstable under the Max-Weight scheduling policy. It seems reasonable, however, that a light-tailed flow that does not conflict with heavy-tailed traffic should be delay stable. Unfortunately, this is not always the case. We demonstrate this by means of simple examples.
 
\par Let us come back to the queueing network of Figure 3. The feasible schedules of this network are $\{1,2\}$ and $\{3\}$, and all queues are served at unit rate, whenever the respective schedules are activated. The rate vector $\lambda=(\lambda_1,\lambda_2,\lambda_3)$ is assumed admissible. The following proposition shows that traffic flow 2 is delay unstable if its rate is sufficiently high.

\medskip\par\textbf{Proposition 1: (Concurring with Heavy Tails)} Consider the single-hop queueing network of Figure 3 under the Max-Weight scheduling policy. If the arriving traffic is admissible and the rates satisfy $\lambda_2>(1+\lambda_1-\lambda_3)/2$, then traffic flow 2 is delay unstable.

\begin{proof}
(Sketch) Let us first give the intuition for the special case, where $\lambda_1=\lambda_3$. Consider sample paths for which a very large file arrives to queue 1; this is a relatively likely event, since traffic flow 1 is heavy-tailed. Queue 3 will build up to the order of magnitude of the large file in queue 1 in order to receive service. Starting from the time slot that the weights of the two schedules become equal, the Max-Weight policy will be draining the weights of the two schedules at the same rate. The period of time until they empty is of the order of magnitude of the large file in queue 1. Now assume that queue 2 stays small throughout this period. If the traffic flows 1 and 3 exhibit their ``average'' behavior, then each feasible schedule will be activated once every two time slots, since $\lambda_1=\lambda_3$. However, if $\lambda_2>1/2$, queue 2 will build up to the order of magnitude of the large file in queue 1, which is a contradiction.

\par The intuition for the more general case is based on the following ``fluid argument'': assume that the arrivals at each queue $f \in \{1,2,3\}$ are a fluid with rate $\lambda_f$. The departures from queue $f$ during periods when all queues are nonempty are also assumed to be a fluid with rate $\mu_f$. The Max-Weight policy has the property of draining the weights of the two feasible schedules at the same rate. Hence, the departure rates are the solution to the following system of linear equations:
\begin{align}
\lambda_1 + \lambda_2 - \mu_1 - \mu_2 &= \lambda_3 - \mu_3 \nonumber \\
\mu_1 + \mu_3 &= 1 \nonumber \\
\mu_1 &= \mu_2. \nonumber
\end{align}

\par The last two equations follow from the facts that Max-Weight is a work-conserving policy, and that queues 1 and 2 are served simultaneously. If the rate at which fluid arrives to queue 2 is greater than the rate at which it departs, i.e.,
\begin{equation}
\lambda_2 > \mu_2 = \frac{1+\lambda_1+\lambda_2-\lambda_3}{3}, \nonumber
\end{equation}
\noindent or, equivalently,
\begin{equation}
\lambda_2 > \frac{1+\lambda_1-\lambda_3}{2}, \nonumber
\end{equation}
\noindent then queue 2 builds up over long periods of time, which, combined with Lemma 1, implies the delay instability of flow 2. A formal proof essentially shows that this fluid model is a faithful approximation of the actual stochastic system (with nonvanishing probability), whenever queue 1 receives a large file; see Appendix 4.
\end{proof}

\medskip\par Proposition 1, as well as Propositions 3 and 4 of the next section, capture a ``rate-dependent phenomenon'' for the propagation of delay instability.

\par We conjecture that a converse to Proposition 1 also holds; namely, that queue 2 is delay stable if the arriving traffic is admissible and $\lambda_2<(1+\lambda_1-\lambda_3)/2$.


\medskip\subsection{Practical Examples and Implications}

\par We illustrate the implications of the results presented so far in the context of specific network topologies, often used to model real-world communication networks.

\medskip\par\textbf{Example 1: (Parallel Queues)} Consider the network of Figure 4, consisting of $n$ parallel queues and a single server. Networks of parallel queues are often used to model wireless uplinks, downlinks, and call centers. Traffic flow 1 is assumed to be heavy-tailed, whereas the other traffic flows are light-tailed. The scheduling constraints of parallel queues require that no two queues can be served simultaneously. The server is allocated according to the Max-Weight scheduling policy, which in this setting is equivalent to ``Serve the Longest Queue.''

\begin{figure}[ht]
\centering
\includegraphics[scale=0.35]{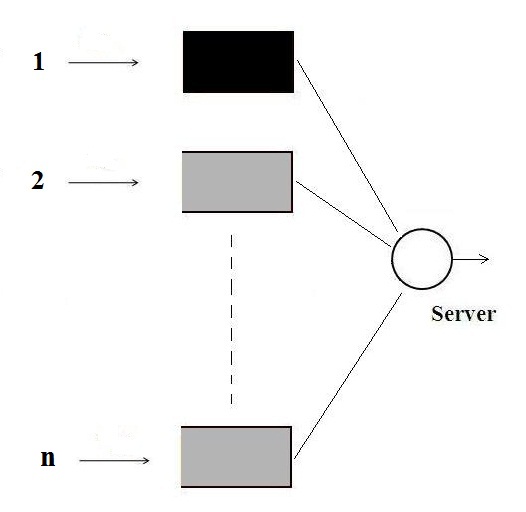}
\caption{Delay instability in parallel queues under Max-Weight scheduling: if traffic flow 1 is heavy tailed (black), then all traffic flows are delay unstable (gray.)}\label{fig:3queues}
\end{figure}

\medskip\par\textbf{Proposition 2:} Consider the system of parallel queues depicted in Figure 4, under the Max-Weight scheduling policy. If traffic flow 1 is heavy-tailed, then all traffic flows are delay unstable.

\begin{proof}
The result follows easily from Theorems 1 and 2.
\end{proof}

\medskip\par\textbf{Example 2: (Input-Queued Switch)} Consider the $2 \times 2$ input-queued switch depicted in Figure 5. Input-queued switches are often used to model internet routers. Traffic flow (1,1) is assumed to be heavy-tailed, whereas all other flows are light-tailed. The scheduling constraints of an input-queued switch require that every feasible schedule has to be a matching between the sets of input and output ports. Thus, the feasible schedules of the network are $\{(1,1),(2,2)\}$ and $\{(1,2),(2,1)\}$. In this setting the Max-Weight scheduling policy activates a matching with the maximum weight.

\begin{figure}[ht]
\centering
\includegraphics[scale=0.6]{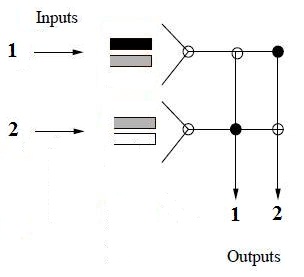}
\caption{Delay instability in a data switch under Max-Weight scheduling: if traffic flow (1,1) is heavy tailed (black), then traffic flows (1,2) and (2,1) are delay unstable (gray.) Traffic flow (2,2) is also delay unstable, if its rate is sufficiently high.}\label{fig:2x2-switch}
\end{figure}

\medskip\par\textbf{Proposition 3:} Consider the $2 \times 2$ input-queued switch depicted in Figure 5, under the Max-Weight scheduling policy. If traffic flow (1,1) is heavy-tailed, then traffic flows (1,1), (1,2), and (2,1) are all delay unstable. If, additionally, $\lambda_{22}>(2+\lambda_{11}-\lambda_{12}-\lambda_{21})/3$, then traffic flow (2,2) is also delay unstable.

\begin{proof}
The first part of the result follows from Theorems 1 and 2. Regarding the second part, we provide the calculations for the associated fluid model, which justify the particular threshold for $\lambda_{22}$: assume that the arrivals at each queue $f \in \{(1,1),(1,2),(2,1),(2,2)\}$ are a fluid with rate $\lambda_f$. The departures from queue $f$ during periods when all queues are nonempty are also assumed to be a fluid with rate $\mu_f$. The Max-Weight policy has the property of draining the weights of the two feasible schedules at the same rate. Hence, the departure rates are the solution to the following system of linear equations:
\begin{align}
\lambda_{11} + \lambda_{22} - \mu_{11} - \mu_{22} &= \lambda_{12} + \lambda_{21} - \mu_{12} - \mu_{21} \nonumber \\
\mu_{11} + \mu_{12} &= 1 \nonumber \\
\mu_{11} &= \mu_{22} \nonumber \\
\mu_{12} &= \mu_{21}. \nonumber
\end{align}

\par The second equation is a consequence of the work-conserving nature of the Max-Weight policy. The last two equations follow from the facts that queue (1,1) is served simultaneously with queue (2,2), and queue (1,2) is served simultaneously with queue (2,1). If the rate at which fluid arrives to queue (2,2) is greater than the rate at which it departs, i.e., if
\begin{equation}
\lambda_{22} > \mu_{22} = \frac{2+\lambda_{11}+\lambda_{22}-\lambda_{12}-\lambda_{21}}{4}, \nonumber
\end{equation}
\noindent or, equivalently, if
\begin{equation}
\lambda_{22} > \frac{2+\lambda_{11}-\lambda_{12}-\lambda_{21}}{3}, \nonumber
\end{equation}
\noindent then queue (2,2) builds up over long periods of time, which, combined with Lemma 1, implies the delay instability of flow (2,2). The proof that the stochastic model follows the fluid model is similar to the proof of Proposition 1 and is omitted.
\end{proof}

\medskip\par\textbf{Example 3: (Wireless Ring)} Consider the wireless ring network of Figure 6. The network consists of 6 nodes, each of which receives traffic that it transmits to its neighboring node in the clockwise direction. Traffic flow 1 is assumed to be heavy-tailed, whereas all other flows are light-tailed. Due to wireless interference, if a link of the network is activated, then the links within two-hop distance must be inactive; this is the so-called two-hop interference model. Thus, the feasible schedules of the network are $\{1,4\}$, $\{2,5\}$, and $\{3,6\}$.

\begin{figure}[ht]
\centering
\includegraphics[scale=0.4]{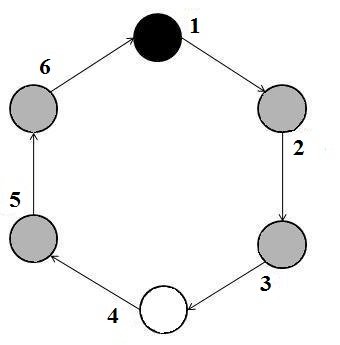}
\caption{Delay instability in a wireless ring network under Max-Weight scheduling: if traffic flow 1 is heavy tailed (black), then traffic flows 2, 3, 5, and 6 are delay unstable (gray.) Traffic flow 4 is also delay unstable, if its rate is sufficiently high.}\label{fig:ringex}
\end{figure}

\medskip\par\textbf{Proposition 4:} Consider the wireless ring network depicted in Figure 6, under the Max-Weight scheduling policy. If traffic flow 1 is heavy-tailed, then traffic flows 1, 2, 3, 5, and 6 are all delay unstable. If, additionally, $\lambda_4>(2+2 \lambda_1-\lambda_2-\lambda_3-\lambda_5-\lambda_6)/4$, then traffic flow 4 is also delay unstable.

\begin{proof}
The first part of the result follows from Theorems 1 and 2. Regarding the second part, we provide the analysis of the associated fluid model: assume that the arrivals at each queue $f \in \{1,2,3,4,5,6\}$ are a fluid with rate $\lambda_f$. The departures from queue $f$ during periods when all queues are nonempty are also assumed to be a fluid with rate $\mu_f$. The Max-Weight policy has the property of draining the weights of the three feasible schedules at the same rate. Hence, the departure rates are the solution to the following system of linear equations:
\begin{align}
\lambda_1 + \lambda_4 - \mu_1 - \mu_4 &= \lambda_2 + \lambda_5 - \mu_2 - \mu_5 \nonumber \\
\lambda_1 + \lambda_4 - \mu_1 - \mu_4 &= \lambda_3 + \lambda_6 - \mu_3 - \mu_6 \nonumber \\
\mu_1 + \mu_2 + \mu_3 &= 1 \nonumber \\
\mu_1 &= \mu_4 \nonumber \\
\mu_2 &= \mu_5 \nonumber \\
\mu_3 &= \mu_6. \nonumber
\end{align}

\par The third equation is a consequence of the work-conserving nature of the Max-Weight policy. The last three equations follow from the facts that queue 1 is served simultaneously with queue 4, and similarly for queues 2 and 5, and queues 3 and 6. If the rate at which fluid arrives to queue 4 is greater than the rate at which it departs, i.e., if
\begin{equation}
\lambda_4 > \mu_4 = \frac{2+2\lambda_1+2\lambda_4-\lambda_2-\lambda_3-\lambda_5-\lambda_6}{6}, \nonumber
\end{equation}
\noindent or, equivalently, if
\begin{equation}
\lambda_4 > \frac{2+2\lambda_1-\lambda_2-\lambda_3-\lambda_5-\lambda_6}{4}, \nonumber
\end{equation}
\noindent then queue 4 builds up over long periods of time, which, combined with Lemma 1, implies the delay instability of flow 4. A detailed proof is omitted for brevity.
\end{proof}


\bigskip\section{Max-Weight-$\alpha$ Scheduling}

\par The results of the previous section suggest that Max-Weight scheduling performs poorly in the presence of heavy-tailed traffic. The reason is that by treating heavy-tailed and light-tailed flows equally, there are very long stretches of time during which heavy-tailed traffic dominates the service. This leads some light-tailed flows to experience very large delays and, eventually, to become delay unstable. Intuitively, by discriminating against heavy-tailed flows one should be able to improve the overall performance of the network, namely to mitigate the propagation of delay instability. One way to do this is by giving preemptive priority to the light-tailed flows. However, priority-based scheduling policies are undesirable because of fairness considerations, and also because they can be unstable in many network settings, e.g., see \cite{KS90,RS92}.

\par Instead, we focus on the Max-Weight-$\alpha$ scheduling policy: given constants $\alpha_f>0$, for all $f \in \{1,\ldots,F\}$, the scheduling vector $S(t)$ belongs to the set:
\begin{equation}
S(t) \ \in \ \arg\max_{(s_f) \in S} \Big\{ \sum_{f=1}^F Q_f^{\alpha_f}(t) \cdot s_f \Big\}. \nonumber
\end{equation}

\par If this set includes multiple feasible schedules, one of them is chosen uniformly at random. By choosing smaller values of the $\alpha$-parameters for heavy-tailed flows and larger values for light-tailed flows, we give a form of partial priority to light-tailed traffic.


\medskip\subsection{The Main Result}

\par Let us start with a preview of the \textbf{main result} of this section: if the $\alpha$-parameters of the Max-Weight-$\alpha$ policy are chosen such that $E[A_f^{\alpha_f+1}(0)]<\infty$, for all $f \in \{1,\ldots,F\}$, then the network is stable and the steady-state queue lengths satisfy:
\begin{equation}
E[Q_f^{\alpha_f}] < \infty, \qquad \forall f \in \{1,\ldots,F\}. \nonumber
\end{equation}

\par An earlier work by Eryilmaz \emph{et al.} has given a similar result for the case of parallel queues with a single server; see Theorem 1 of \cite{ESP05}. In this paper we extend their result to a general single-hop network setting. Moreover, we provide an explicit upper bound to the sum of the $\alpha$-moments of the steady-state queue lengths. Before we do that we need the following definition.

\medskip\par\textbf{Definition 7: (Covering Number of Feasible Schedules)} The covering number $k^*$ of the set of feasible schedules is defined as the smallest number $k$ for which there exist $s^1,\ldots,s^k \in S$ with $\bigcup_{i=1}^k s^i = \{1,\ldots,F\}$.

\medskip\par Notice that the quantity $k^*$ is a structural property of the queueing network, and is not related to the scheduling policy or the statistics of the arriving traffic: it is the minimum number of time slots required to serve at least one packet from each flow.

\medskip\par\textbf{Theorem 3: (Max-Weight-$\alpha$ Scheduling)} Consider the single-hop queueing network described in Section 2 under the Max-Weight-$\alpha$ scheduling policy. Let the intensity of the arriving traffic be $\rho<1$. If $E[A_f^{\alpha_f+1}(0)]<\infty$, for all $f \in \{1,\ldots,F\}$, then the queueing network is stable and the steady-state queue lengths satisfy:
\begin{equation}
\sum_{f=1}^F E[Q_f^{\alpha_f}] \leq \sum_{f=1}^F H \Big( \rho,k^*,\alpha_f,E[A_f^{\alpha_f+1}(0)] \Big), \nonumber
\end{equation}
\noindent where
\begin{equation}
H \Big( \rho,k^*,\alpha_f,E[A_f^{\alpha_f+1}(0)] \Big) = \left\{ \begin{array}{ll} \frac{2k^*}{1-\rho} \cdot \Big( E[A_f^{\alpha_f+1}(0)]+1 \Big), & \alpha_f \leq 1, \\
\Big( \frac{2k^*}{1-\rho} \Big)^{\alpha_f} \cdot K^{\alpha_f}  + \frac{2k^*}{1-\rho} \cdot K, & \alpha_f > 1,
\end{array} \right. \nonumber
\end{equation}

\noindent and $K=2^{\alpha_f-1} \cdot \alpha_f \cdot \Big( E[A_f^{\alpha_f+1}(0)]+1 \Big)$.

\medskip\begin{proof}
(Sketch) Consider the single-hop queueing network of Section 2 under the Max-Weight-$\alpha$ scheduling policy. It can be verified that the sequence $\{Q(t);\ t \in Z_+\}$ is a time-homogeneous, irreducible, and aperiodic Markov chain on the countable state-space $Z_+^F$. The fact that this Markov chain is also positive recurrent, and the related moment bound, are based on drift analysis of the Lyapunov function
\begin{equation}
V(Q(t)) = \sum_{f=1}^F \frac{1}{\alpha_f+1} \cdot Q_f^{\alpha_f+1}(t), \nonumber
\end{equation}
\noindent and use of the Foster-Lyapunov stability criterion. This implies that $\{Q(t);\ t \in Z_+\}$ converges in distribution, and its limiting distribution does not depend on $Q(0)$. Based on this, it can be verified that the sequence $\{D(k);\ k \in N\}$ is a (possibly delayed) aperiodic and positive recurrent regenerative process. Hence, it also converges in distribution, and its limiting distribution does not depend on $Q(0)$. For a formal proof see Appendix 5.
\end{proof}


\medskip\subsection{Traffic Burstiness and Delay Stability}

\par A first corollary of Theorem 3 relates to the delay stability of light-tailed flows.

\medskip\par\textbf{Corollary 1: (Delay Stability under Max-Weight-$\alpha$)} Consider the single-hop queueing network described in Section 2 under the Max-Weight-$\alpha$ scheduling policy. If the $\alpha$-parameters of all light-tailed flows are equal to 1, and the $\alpha$-parameters of heavy-tailed flows are sufficiently small, then all light-tailed flows are delay stable.

\begin{proof}
With the particular choice of $\alpha$-parameters, Theorem 3 guarantees that the expected steady-state queue length of all light-tailed flows is finite. Lemma 1 relates this result to delay stability.
\end{proof}

\medskip\par Combining this with Theorem 1, we conclude that when its $\alpha$-parameters are chosen suitably, \textbf{the Max-Weight-$\alpha$ policy delay-stabilizes a traffic flow, whenever this is possible}.

\par Max-Weight-$\alpha$ turns out to perform well in terms of another criterion too. Theorem 3 implies that by choosing the $\alpha$-parameters such that $E[A_f^{\alpha_f+1}(0)]<\infty$, for all $f \in \{1,\ldots,F\}$, the steady-state queue length moment $E[Q_f^{\alpha_f}]$ is finite, for all $f \in \{1,\ldots,F\}$. The following proposition suggests that this is the best we can do under any regenerative scheduling policy.

\medskip\par\textbf{Proposition 5:} Consider the single-hop queueing network described in Section 2 under a regenerative scheduling policy. Then,
\begin{equation}
E[A_f^{c+1}(0)]=\infty \ \Longrightarrow \ E[Q_f^c]=\infty, \qquad \forall f \in \{1,\ldots,F\}. \nonumber
\end{equation} 

\begin{proof}
This result is well-known in the context of a M/G/1 queue, e.g., see Section 3.2 of \cite{BZ07}. It can be proved similarly to Theorem 1.
\end{proof}

\medskip\par Thus, when its $\alpha$-parameters are chosen suitably, \textbf{the Max-Weight-$\alpha$ policy guarantees the finiteness of the highest possible moments of steady-state queue lengths}.


\medskip\subsection{Scaling Results under Light-Tailed Traffic}

\par Although this paper focuses on heavy-tailed traffic and its consequences, some implications of Theorem 3 are of general interest. In this section we assume that all traffic flows in the network are light-tailed, and analyze how the sum of the $\alpha$-moments of steady-state queue lengths scales with traffic intensity and the size of the network.

\medskip\par\textbf{Corollary 2: (Scaling with Traffic Intensity)} Let us fix a single-hop queueing network and constants $\alpha \geq 1$ and $B>0$. The Max-Weight-$\alpha$ scheduling policy is applied with $\alpha_f = \alpha$, for all $f \in \{1,\ldots,F\}$. Assume that the traffic arriving to the network is admissible, and that the $(\alpha+1)$-moments of all traffic flows are bounded from above by $B$. Then,
\begin{equation}
\sum_{f=1}^F E[Q_f^{\alpha}] \leq \frac{M(k^*,\alpha,B)}{(1-\rho)^{\alpha}}, \nonumber
\end{equation}
\noindent where $M(k^*,\alpha,B)$ is a constant that depends only on $k^*$, $\alpha$, and $B$. Moreover, under any stabilizing scheduling policy
\begin{equation}
\sum_{f=1}^F E[Q_f^{\alpha}] \geq \frac{M'(\alpha)}{(1-\rho)^{\alpha}}, \nonumber
\end{equation}
\noindent where $M'(\alpha)$ is a constant that depends only on $\alpha$.

\begin{proof} 
If $\alpha_f = \alpha \geq 1$, for all $f \in \{1,\ldots,F\}$, then Theorem 3 implies that:
\begin{equation}
\sum_{f=1}^F E[Q_f^{\alpha}] \leq \frac{M(k^*,\alpha,B)}{(1-\rho)^{\alpha}}, \nonumber
\end{equation}
\noindent where $M(k^*,\alpha,B)$ is a constant that depends only on $k^*$, $\alpha$, and $B$.

\par On the other hand, Theorem 2.1 of \cite{SW08} implies that under any stabilizing scheduling policy there exists an absolute constant $\tilde{M}$, such that
\begin{equation}
\sum_{f=1}^F E[Q_f] \geq \frac{\tilde{M}}{(1-\rho)}. \nonumber
\end{equation}

\noindent Utilizing Jensen's inequality, we have:
\begin{align}
\sum_{f=1}^F E[Q_f^{\alpha}] &\geq \sum_{f=1}^F (E[Q_f])^{\alpha} \nonumber \\
&\geq \frac{1}{F^{\alpha}} \Big( \sum_{f=1}^F E[Q_f] \Big)^{\alpha}. \nonumber
\end{align}

\noindent Consequently, there exists a constant $M'(\alpha)$ that depends only on $\alpha$, such that
\begin{equation}
\sum_{f=1}^F E[Q_f^{\alpha}] \geq \frac{M'(\alpha)}{(1-\rho)^{\alpha}}, \nonumber
\end{equation}
\noindent under any stabilizing scheduling policy.
\end{proof}

\medskip\par Similar scaling results appear in queueing theory, mostly in the context of single-server queues, e.g., see Chapter 3 of \cite{H06}. More recently, results of this flavor have been shown for particular queueing networks, such as input-queued switches \cite{STZ11,SW08}. All the related work, though, concerns the scaling of first moments. Corollary 2 gives the precise scaling of higher order steady-state queue length moments with traffic intensity, and shows that Max-Weight-$\alpha$ achieves the \textbf{optimal scaling}.

\par We now turn our attention to the performance of the Max-Weight scheduling policy under Bernoulli traffic, i.e., when each of the arrival processes $\{A_f(t);\ t \in Z_+\}$ is an independent Bernoulli process with parameter $\lambda_f>0$.

\par We denote by $S_{\max}$ the maximum number of traffic flows that any feasible schedule $s \in S$ can serve.

\medskip\par\textbf{Corollary 3: (Scaling under Bernoulli Traffic)} Consider the single-hop queueing network described in Section 2 under the Max-Weight scheduling policy. Assume that the traffic arriving to the network is Bernoulli, with traffic intensity $\rho<1$. Then,
\begin{equation}
\sum_{f=1}^F E[Q_f] \leq 2 \cdot k^* \cdot S_{\max} \cdot \Big( \frac{1+\rho}{1-\rho} \Big). \nonumber 
\end{equation}

\begin{proof}
If all traffic flows are light-tailed and all the $\alpha$-parameters are equal to one, a more careful accounting in the proof of Theorem 3 provides the following tighter upper bound:
\begin{equation}
\sum_{f=1}^F E[Q_f] \leq \frac{2k^*}{1-\rho} \cdot \Big( S_{\max} + \sum_{f=1}^F E[A_f^2(0)] \Big). \nonumber 
\end{equation}

\par If the traffic arriving to the network is Bernoulli, then $E[A_f^2(0)]=\lambda_f$, for all $f \in \{1,\ldots,F\}$. Moreover, the fact that the arriving traffic has intensity $\rho$, implies the existence of nonnegative real numbers $\zeta_s$, for $s \in S$, such that:
\begin{equation}
\lambda_f \leq \sum_{s \in S} \zeta_s \cdot s_f, \qquad \forall f \in \{1,\ldots,F\}, \nonumber
\end{equation}
\noindent and
\begin{equation}
\sum_{f=1}^F \zeta_s = \rho. \nonumber
\end{equation}

\noindent Consequently,
\begin{align}
\sum_{f=1}^F E[A_f^2(0)] &= \sum_{f=1}^F \lambda_f \nonumber \\
&\leq \sum_{f=1}^F \sum_{s \in S} \zeta_s \cdot s_f \nonumber \\
&= \sum_{s \in S} \zeta_s \cdot \sum_{f=1}^F \cdot s_f \nonumber \\
&\leq \sum_{s \in S} \zeta_s \cdot S_{\max} \nonumber \\
&= \rho \cdot S_{\max}, \nonumber
\end{align}
\noindent and the result follows.
\end{proof}

\medskip\par\textbf{Example 4: ($n$ Parallel Queues)} Consider a single-server system with $n$ parallel queues. The arriving traffic is assumed to be Bernoulli, with traffic intensity $\rho<1$. In this case $k^*=n$ and $S_{\max}=1$. Corollary 3 implies that under the Max-Weight scheduling policy, the sum of the steady-state queue lengths is bounded from above by:
\begin{equation}
\sum_{i=1}^n E[Q_i] \leq \frac{ 4 n}{1-\rho}. \nonumber
\end{equation}

\medskip\par The total queue length of a system of parallel queues under a work-conserving scheduling policy evolves like a $Geo^{[B]}/D/1$ queue, from which we infer that $\sum_{i=1}^n E[Q_i] = \Theta \Big( \frac{1}{1-\rho} \Big)$. So, in the context of parallel queues, the scaling provided by Corollary 3 is tight with respect to the traffic intensity, but not necessarily tight with respect to the size of the network.

\medskip\par\textbf{Example 5: ($n \times n$ Input-Queued Switch)} Consider a $n \times n$ input-queued switch. The arriving traffic is assumed to be Bernoulli, with traffic intensity $\rho<1$. In this case $k^*=n$ and $S_{\max}=n$. Corollary 3 implies that under the Max-Weight scheduling policy, the sum of the steady-state queue lengths is bounded from above by:
\begin{equation}
\sum_{i=1}^n \sum_{j=1}^n E[Q_{ij}] \leq \frac{4 n^2}{1-\rho}. \nonumber
\end{equation}

\medskip\par In the context of input-queued switches, the joint scaling provided by Corollary 3, in terms of both the traffic intensity and the size of the network, is the tightest currently known. However, it should be noted that the correct scaling as $n\to\infty$ and $\rho \to 1$ is an open problem; see \cite{STZ11}.

\medskip\par\textbf{Example 6: ($n \times n$ Grid)} Consider a single-hop queueing network in a $n \times n$ grid topology, under the one-hop interference model. The arriving traffic is assumed to be Bernoulli, with traffic intensity $\rho<1$. In this case $k^* \leq 4$ and $S_{\max} \leq n^2/2$. Corollary 3 implies that under the Max-Weight scheduling policy, the sum of the steady-state queue lengths is bounded from above by:
\begin{equation}
\sum_{i=1}^n \sum_{j=1}^n E[Q_{ij}] \leq \frac{8 n^2}{1-\rho}. \nonumber
\end{equation}


\bigskip\section{Discussion}

\par The main conclusion of this paper is that the celebrated Max-Weight scheduling policy performs poorly in the presence of heavy-tailed traffic. More specifically, our findings show that the phenomenon of delay instability not only arises, but can propagate to a significant part of the network. This is somewhat surprising, since Max-Weight is known to perform very well in the presence of light-tailed traffic, at least in single-hop queueing networks.

\par Another important conclusion is that the Max-Weight-$\alpha$ scheduling policy can be used to alleviate the effects of heavy-tailed traffic, and is even order optimal, if its $\alpha$-parameters are chosen suitably. However, for Max-Weight-$\alpha$ to perform well, accurate knowledge of the tail coefficients of all traffic flows is required. If the $\alpha$-parameters are not chosen appropriately, then in light of Proposition 5, this policy may also perform poorly.

\par Of particular interest is the study of networks with time-varying channel state. In this class of models there exists an underlying state of the network which evolves in time, and the transmission rates of the links are given by a function of the state. Under certain conditions on the channel state evolution, it can be verified that Theorems 1-3 carry over with minimal changes to this more general setting.

\par An important direction for future research is to consider queueing networks with correlated traffic. The IID assumption that we made here facilitates the analysis and offers valuable insights, but is clearly restrictive. As alluded to earlier, evidence suggests that traffic in real-world networks exhibits strong correlations, and phenomena such as self-similarity and long-range dependence arise. Concrete results in this direction would be of great theoretical and practical interest.



\bigskip\section*{Appendix 1 - Background Material}

\subsection*{1.1 \ \ BASTA, Little's Law, and Delay Stability}

\par In this section we give the ``steady-state versions'' of two important results in queueing theory, the Bernoulli Arrivals See Time Averages property and Little's Law, which we later use to prove Lemma 1.

\par Consider the single-hop queueing network described in Section 2. Let $\tau_{f,k}$ be the random time slot of the arrival of the $k^{th}$ file to queue $f$, $k \in N,\ f \in \{1,\ldots,F\}$. We assign two marks to this file: i) the vector of queue lengths upon its arrival $Q^{c(f)}(k)=(Q_g(\tau_{f,k});\ g=1,\ldots,F)$; and ii) its end-to-end delay $D_f(k)$.

\par Under a regenerative scheduling policy, and for a given $f \in \{1,\ldots,F\}$, the vector-valued sequences $\{Q^{c(f)}(k);\ k \in N\}$, as well as the sequence $\{Q(t);\ t \in Z_+\}$, are (possibly delayed) aperiodic and positive recurrent regenerative processes. Therefore, they converge in distribution, and their limiting distributions do not depend on $Q(0)$; see \cite{SW93}. We denote by $Q^{c(f)}=(Q_g^{c(f)}),\ g=1,\ldots,F\}$ and $Q=(Q_g;\ g=1,\ldots,F)$ generic random vectors distributed according to these limiting distributions.

\par The arrival of files at queue $f$ constitutes a Bernoulli process with parameter $p_f=P(A_f(0)>0),\ f \in \{1,\ldots,F\}$, since all arrival processes are IID. The Bernoulli Arrivals See Time Averages (BASTA) property relates the limiting distributions $Q^{c(f)}$ and $Q$.

\medskip\par\textbf{Theorem 4: (BASTA)} Consider the single-hop queueing network described in Section 2 under a regenerative scheduling policy. Then,
\begin{equation}
Q^{c(f)} \stackrel{d}{=} Q, \qquad \forall f \in \{1,\ldots,F\}, \nonumber
\end{equation}
\noindent where $\stackrel{d}{=}$ denotes equality in distribution.

\begin{proof}
Fix a queue $f \in \{1,\ldots,F\}$ and consider the random variables:
\begin{equation}
U_T = \frac{1}{T} \sum_{t=0}^{T-1} 1_{\{Q(t) \leq B\}}, \nonumber
\end{equation}

\noindent and
\begin{equation}
V_K = \frac{1}{K} \sum_{k=1}^{K} 1_{\{Q^{c(f)}(k) \leq B\}}, \nonumber
\end{equation}

\noindent where $T,K \in N$, and $B \in Z_+^F$. The conditions of Theorem 3 in \cite{MMW89} are satisfied, and we have:
\begin{equation}
\lim_{K \to \infty} V_K = \lim_{T \to \infty} U_T \qquad \mbox{w.p.1}. \nonumber
\end{equation}

\par Under a regenerative scheduling policy the sequences $\{1_{\{Q(t) \leq B\}};\ t \in Z_+\}$ and $\{1_{\{Q^{c(f)}(k) \leq B\}};\ k \in N\}$ are (possibly delayed) positive recurrent regenerative processes, which are also uniformly bounded by one. Then, the Ergodic theorem for regenerative processes implies that
\begin{equation}
\lim_{T \to \infty} U_T = \lim_{T \to \infty} \frac{1}{T} \sum_{t=0}^{T-1} 1_{\{Q(t) \leq B\}} = P(Q \leq B) \qquad \mbox{w.p.1}, \nonumber \\
\end{equation}

\noindent and
\begin{equation}
\lim_{K \to \infty} V_K = \lim_{K \to \infty} \frac{1}{K} \sum_{k=1}^{K} 1_{\{Q^{c(f)}(k) \leq B\}} = P(Q^{c(f)} \leq B) \qquad \mbox{w.p.1}; \nonumber
\end{equation}

\noindent see \cite{SW93}. Consequently,
\begin{equation}
P(Q \leq B) = P(Q^{c(f)} \leq B), \qquad \forall B \in Z_+^F, \nonumber
\end{equation}
\noindent and the result follows.
\end{proof}

\medskip\par Now let $L_f(t)$ be the number of files in queue $f$ at time slot $t$, either queued up or in service. Under a regenerative scheduling policy, the sequences $\{L_f(t);\ t \in Z_+\}$ and $\{D_f(k);\ k \in N\},\ f \in \{1,\ldots,F\}$ are (possibly delayed) aperiodic and positive recurrent regenerative processes. Hence, they converge in distribution, and their limiting distributions do not depend on $Q(0)$; see \cite{SW93}. We denote by $L_f$ and $D_f$ generic random variables distributed according to these limiting distributions. Little's Law relates the expected values of these limiting distributions.

\medskip\par\textbf{Theorem 5: (Little's Law)} Consider the single-hop queueing network described in Section 2 under a regenerative scheduling policy. Then,
\begin{equation}
E[L_f] = p_f \cdot E[D_f], \qquad \forall f \in \{1,\ldots,F\}. \nonumber
\end{equation}
\noindent Furthermore, this is true even if these expectations are infinite.

\begin{proof}
\par First, we establish Little's Law for the case of finite expectations. Fix a queue $f \in \{1,\ldots,F\}$, and assume that $E[L_f]$ is finite. We call the aggregate length of queue $f$ during a regeneration cycle, and write $L^f_{agg}$, the random variable
\begin{equation} 
L^f_{agg} = \sum_{t=\tau_0}^{\tau_1-1} L_f(t), \nonumber
\end{equation}
\noindent where $\tau_0$ and $\tau_1$ represent the first two (or, in general, two consecutive) regeneration epochs of the network. 

\par Initially, we prove by contradiction that $E[L^f_{agg}]$ is finite. Suppose that $E[L^f_{agg}]$ is infinite. Using a truncation argument, similar to the one in Lemma 4 of Appendix 1.3, it can be shown that $E[L_f]$ is also infinite. This contradicts our assumption that $E[L_f]$ is finite. Hence, $E[L^f_{agg}]$ is finite. 

\par The sequence $\{L_f(t);\ t \in Z_+\}$ is a (possibly delayed) positive recurrent regenerative process. Combined with the fact that $E[L^f_{agg}]$ is finite, the Ergodic theorem for regenerative processes implies that
\begin{equation}
\lim_{T \to \infty} \frac{1}{T} \sum_{t=0}^{T-1} L_f(t) = E[L_f] \qquad \mbox{w.p.1}; \nonumber
\end{equation}

\noindent see \cite{SW93}. Moreover, since the network is stable under a regenerative scheduling policy,
\begin{equation}
\lim_{k \to \infty} \frac{D_f(k)}{k} = 0 \qquad \mbox{w.p.1}; \nonumber
\end{equation} 
see Theorem 2b of \cite{GW86}. The sequence $\{D_f(k);\ k \in N\}$ is also a (possibly delayed) positive recurrent regenerative process. Then, the Ergodic theorem for regenerative processes and Theorem 2e of \cite{GW86} imply that
\begin{align}
\lim_{K \to \infty} \frac{1}{K} \sum_{k=1}^{K} D_f(k) = E[D_f] \qquad \mbox{w.p.1}, \nonumber
\end{align}

\noindent and
\begin{equation}
E[L_f] = p_f \cdot E[D_f]. \nonumber
\end{equation}

\par To summarize, starting with the assumption that $E[L_f]$ is finite, we showed that $E[L_f] = p_f \cdot E[D_f]$. The same can be shown if we start with the assumption $E[D_f]$ is finite, and work similarly. Consequently,
\begin{equation} 
E[L_f]<\infty \ \Longleftrightarrow \ E[D_f]<\infty, \nonumber
\end{equation}
\noindent which implies that Little's Law holds even if the implicated expectations are infinite.
\end{proof}

\medskip\par We now re-state and prove Lemma 1.

\medskip\par\textbf{Lemma 1:} Consider the single-hop queueing network described in Section 2 under a regenerative scheduling policy. Then,
\begin{equation}
E[Q_f]<\infty \ \Longleftrightarrow \ E[D_f]<\infty, \qquad \forall f \in \{1,\ldots,F\}. \nonumber
\end{equation}

\begin{proof}
\par Let us start with the implication 
\begin{equation}
E[Q_f]<\infty \ \Longrightarrow \ E[D_f]<\infty, \qquad \forall f \in \{1,\ldots,F\}. \nonumber
\end{equation}

\noindent Assume that $E[Q_f]$ is finite, for some $f \in \{1,\ldots,F\}$. Since every file has at least one packet,
\begin{equation}
P(Q_f(t)>B) \geq P(L_f(t)>B), \qquad \forall t \in Z_+, \qquad \forall B \in Z_+. \nonumber
\end{equation}

\par We have argued that under a regenerative scheduling policy, the sequences $\{Q_f(t);\ t \in Z_+\}$ and $\{L_f(t);\ t \in Z_+\}$ converge in distribution. So, taking the limit as $t$ goes to infinity, we have:
\begin{equation}
P(Q_f>B) \geq P(L_f>B), \qquad \forall B \in Z_+, \nonumber
\end{equation}

\noindent which, in turn, implies that
\begin{equation}
E[Q_f] \geq E[L_f]. \nonumber
\end{equation}

\noindent Combining this inequality with Little's Law and the assumption that $E[Q_f]$ is finite, we conclude that
\begin{equation}
E[D_f]<\infty. \nonumber
\end{equation}

\par Let us now prove the implication 
\begin{equation}
E[Q_f]=\infty \ \Longrightarrow \ E[D_f]=\infty, \qquad \forall f \in \{1,\ldots,F\}. \nonumber
\end{equation}

\noindent Assume that $E[Q_f]$ is infinite, $f \in \{1,\ldots,F\}$. The end-to-end delay of a file is bounded from below by the length of the respective queue upon its arrival, since the service discipline within each queue is ``First Come, First Served.'' So,
\begin{equation}
P(D_f(k)>B) \geq P(Q_f(\tau_{f,k})>B), \qquad \forall k \in N, \qquad \forall B \in Z_+. \nonumber
\end{equation}

\par We have argued that under a regenerative scheduling policy, the sequences $\{D_f(k);\ k \in N\}$ and $\{Q_f(\tau_{f,k});\ k \in N\}$ converge in distribution. So, taking the limit as $k$ goes to infinity, we have 
\begin{equation}
P(D_f>B) \geq P(Q^{c(f)}_f>B), \qquad \forall B \in Z_+. \nonumber
\end{equation}

\noindent Combining this with the BASTA property,
\begin{equation}
P(D_f>B) \geq P(Q_f>B), \qquad \forall B \in Z_+, \nonumber
\end{equation}

\noindent which results in 
\begin{equation}
E[D_f] \geq E[Q_f]. \nonumber
\end{equation}

\noindent Finally, the assumption that $E[Q_f]$ is infinite implies that
\begin{equation}
E[D_f]=\infty. \nonumber
\end{equation}
\end{proof}


\bigskip\subsection*{1.2 \ \ ``Average'' Behavior of an IID Sequence}

\par The following result is a well-known corollary of the Strong Law of Large Numbers. We provide a proof for completeness.

\medskip\par\textbf{Lemma 3:} Consider a sequence of IID random variables $\{B(\tau); \ \tau \in N\}$, taking values in $Z_+$, with finite rate $\lambda = E[B(1)]>0$. For any given $\epsilon>0$, there exists a constant $\delta>0$, such that
\begin{equation}
P \Big( \Big\{(\lambda-\epsilon)t-\delta \leq \sum_{\tau=1}^{t} B(\tau) \leq (\lambda+\epsilon)t+\delta \Big\}, \ \forall t \in N \Big)>0. \nonumber
\end{equation}
\begin{proof}
We define an event $C_m$ by
$$C_m=\Big\{ \, \Big| \frac{1}{t} \sum_{\tau=1}^t B(\tau) -\lambda \Big|
\leq \epsilon,\ \ \forall\ t\geq m\Big\}.$$
By the strong law of large numbers, $P(\cup_{m\geq 1} C_m)=1$. Because the sequence of events $C_m$ is nondecreasing,  the continuity property of probabilities implies that $\lim_{m\to\infty}P(C_m)=1$. Let us therefore fix some $T$ such that $P(C_T)>1/2$.

Let us consider the event 
$$D=\Big\{\, 0\leq \sum_{\tau=1}^T B(\tau) \leq \delta\Big\}.$$
We choose $\delta$ large enough so that $P(D)>1/2$ and $\delta\geq\lambda T$. Note that
$$P(C_T \cap D)\geq P(C_T)+P(D)-1 >\frac{1}{2}+\frac{1}{2}-1=0.$$
Note also that when both $C_T$ and $D$ occur, then
$$\Big| \sum_{\tau=1}^t B(\tau) -\lambda t\Big| \leq \epsilon t +\delta,\qquad \forall\ t,$$
so that the latter event has positive probability, 
which is the desired result follows.
\end{proof}


\bigskip\subsection*{1.3 \ \  Truncated Rewards}

\par Consider the single-hop queueing network described in Section 2 under a regenerative scheduling policy. By definition, there exists a sequence of stopping times $\{\tau_n;\ n \in Z_+\}$, which constitutes a (possibly delayed) renewal process, i.e., the sequence $\{\tau_{n+1}-\tau_n;\ n \in Z_+\}$ is IID. Moreover, the lattice distribution of cycle lengths has span equal to one and finite expectation. 

\par For $t \in Z_+$, let $R(t)$ be an instantaneous reward on this renewal process, which is assumed to be an arbitrary function of $Q(t)$. We define the truncated reward as $R^M(t) = \min\{R(t),M\}$, where $M$ is a positive integer. Under a regenerative scheduling policy, the sequences $\{R(t);\ t \in Z_+\}$ and $\{R^M(t);\ t \in Z_+\}$ are (possibly delayed) aperiodic and positive recurrent regenerative processes. Consequently, they converge in distribution, and their limiting distributions do not depend on $Q(0)$; see \cite{SW93}. Let $R$ and $R^M$ be generic random variables distributed according to these limiting distributions. We denote by $R_{agg}$ the aggregate reward, i.e., the reward accumulated over a regeneration cycle. Similarly, $R^M_{agg}$ represents the truncated aggregate reward. 

\medskip\par\textbf{Lemma 4:} Consider the single-hop queueing network described in Section 2 under a regenerative scheduling policy. Suppose that there exists a random variable $Y$ with infinite expectation, and a nondecreasing function $f(\cdot)$, such that $\lim_{M\to\infty}f(M)=\infty$, and
\begin{equation}
E[\min\{Y,f(M)\}] \leq E[R^M_{agg}]. \tag{1}
\end{equation}

\noindent Then,
\begin{equation}
E[R]=\infty. \nonumber
\end{equation}

\begin{proof}
\par By definition, cycle lengths have finite expectation, and $E[R^M_{agg}]$ is bounded from above by $M \cdot E[\tau_1-\tau_0]$. Then, the Renewal Reward theorem implies that
\begin{equation}
\frac{E[R^M_{agg}]}{E[\tau_1-\tau_0]} = \lim_{T \to \infty} \frac{1}{T} \sum_{t=0}^{T-1} R^M(t), \qquad \mbox{w.p.1}; \tag{2}
\end{equation}

\noindent see Section 3.4 of \cite{G96}. The sequence $\{R^M(t);\ t \in Z_+\}$ is a (possibly delayed) positive recurrent regenerative process, which is also uniformly bounded by $M$. Then, the Ergodic theorem for regenerative processes implies that
\begin{equation}
\lim_{T \to \infty} \frac{1}{T} \sum_{t=0}^{T-1} R^M(t) = \lim_{T \to \infty} \frac{1}{T} \sum_{t=0}^{T-1} \min\{R(t),M\} = E[\min\{R,M\}], \qquad \mbox{w.p.1}; \tag{3}
\end{equation}
see \cite{SW93}. Eqs.\ (1)-(3) give:
\begin{equation}
\frac{E[\min\{Y,f(M)\}]}{E[\tau_1-\tau_0]} \leq E[\min\{R,M\}]. \nonumber
\end{equation}
By taking the limit as $M$ goes to infinity on both sides, and using the Monotone Convergence theorem, we obtain
\begin{equation}
\frac{E[Y]}{E[\tau_1-\tau_0]} \leq E[R]; \nonumber
\end{equation}
see Section 5.3 of \cite{WI91}. Finally, the fact that $Y$ has infinite expectation implies that
\begin{equation}
E[R]=\infty. \nonumber
\end{equation}
\end{proof}


\bigskip\section*{Appendix 2 - Proof of Theorem 1}

\par Consider a heavy-tailed traffic flow $h \in \{1,\ldots,F\}$. We will show that under any regenerative scheduling policy:
\begin{equation}
E[Q_h]=\infty. \nonumber
\end{equation}
\noindent Combined with Lemma 1, this will imply that traffic flow $h$ is delay unstable.

\par Consider a fictitious queue, denoted by $\tilde{h}$, which has exactly the same arrivals and initial length as queue $h$, but is served at unit rate whenever nonempty. We denote by $Q_{\tilde{h}}(t)$ the length of queue $\tilde{h}$ at time slot $t$. Since the arriving traffic is assumed admissible, the queue length process $\{Q_{\tilde{h}}(t);\ t \in Z_+\}$ converges to a limiting distribution $Q_{\tilde{h}}$.

\par An easy, inductive argument can show that under a regenerative scheduling policy, the length of queue $h$ dominates the length of queue $\tilde{h}$ at all time slots. This implies that
\begin{equation}
P(Q_h(t)>B) \geq P(Q_{\tilde{h}}(t)>B), \qquad \forall t \in Z_+, \quad \forall B \in Z_+. \nonumber
\end{equation}
\noindent Taking the limit as $t$ goes to infinity, and using the fact that both queue length processes converge in distribution, we have:
\begin{equation}
P(Q_h>B) \geq P(Q_{\tilde{h}}>B), \qquad \forall B \in Z_+. \nonumber
\end{equation}

\noindent In order to prove the desired result, it suffices to show that 
\begin{equation}
E[Q_{\tilde{h}}]=\infty. \nonumber
\end{equation}

\par The time slots that initiate busy periods of queue $\tilde{h}$ constitute regeneration epochs. Denote by $X_i$ the length of the $i^{th}$ cycle. The random variables $\{X_i;\ i \in N\}$ are IID copies of some nonnegative random variable $X$, with finite first moment; this is because the length of queue $\tilde{h}$ is a positive recurrent Markov chain, and the empty state is recurrent.

\par We define an instantaneous reward on this renewal process:
\begin{equation}
R^M(t) = \min\{Q_{\tilde{h}}(t),M\}, \qquad \forall t \in Z_+, \nonumber
\end{equation}
where $M$ is some finite integer.

\par Without loss of generality, assume that a busy period starts at time slot 0, and let $B$ be the random size of the file that initiates it. Since queue $\tilde{h}$ is served at unit rate, its length is at least $B/2$ packets over a time period of length at least $B/2$ time slots. This implies that the aggregate reward $R^M_{agg}$, i.e., the reward accumulated over a renewal period, is bounded from below by
\begin{align}
R^M_{agg} &\geq \frac{B}{2} \cdot \min \Big\{ \frac{B}{2} ,M \Big\} \nonumber \\
&\geq \min \Big\{ \frac{B^2}{4} ,M^2 \Big\}. \nonumber
\end{align}
Consequently, the expected aggregate reward is bounded from below by 
\begin{align}
E[R^M_{agg}] &\geq \sum_{B=1}^{\infty} \min \Big\{ \frac{B^2}{4} ,M^2 \Big\} \cdot P(A_h(0)=B) \nonumber \\
&= \sum_{B=0}^{\infty} \min \Big\{ \frac{B^2}{4} ,M^2 \Big\} \cdot P(A_h(0)=B) \nonumber \\
&= E \Big[\min \Big\{ \frac{A_h^2(0)}{4} ,M^2 \Big\} \Big]. \nonumber
\end{align}
Then, Lemma 4 (see Appendix 1.3) applied to $Y=(1/4) A^2_h(0)$, implies that $E[Q_{\tilde{h}}]=\infty$. This, in turn, gives:
\begin{equation}
E[Q_h]=\infty. \nonumber
\end{equation}


\bigskip\section*{Appendix 3 - Proof of Theorem 2}

\par Consider a heavy-tailed traffic flow $h$, and a light-tailed flow $l$ that conflicts with $h$. We will show that for admissible traffic flow rates and under the Max-Weight scheduling policy:
\begin{equation}
E[Q_l]=\infty. \nonumber
\end{equation}
\noindent Combined with Lemma 1, this will imply that traffic flow $l$ is delay unstable.

\par The time slots that initiate busy periods of the network constitute regeneration epochs. Denote by $X_i$ the length of the $i^{th}$ cycle. The random variables $\{X_i; \ i \in N\}$ can be viewed as IID copies of some nonnegative random variable $X$, with finite first moment; this is because the network is stable under the Max-Weight policy and the empty state is recurrent. 

\par We define an instantaneous reward on this renewal process:
\begin{equation}
R^M(t) = \min\{Q_l(t),M\}, \qquad \forall t \in Z_+, \nonumber
\end{equation}
\noindent where $M$ is a positive integer.

\par Without loss of generality, assume that a renewal period of the network starts at time slot 0. Consider the set of sample paths where at time slot 0, queue $h$ receives a file of size $B$ packets, and all other queues receive no traffic; we denote this set of sample paths by $H(B)$. Since the arrival processes of different traffic flows are mutually independent, $P(H(B))= P(A_h(0)=B) \cdot \prod_{g \neq h} P(A_g(0)=0)$. For sample paths in $H(B)$, denote by $T_B$ the first time slot when the length of queue $h$ becomes less than or equal to the sum of the lengths of all other queues:
\begin{equation}
T_B = \min \Big\{ t>0 \ \Big| \ \sum_{g \neq h} Q_g(t) \geq Q_h(t) \Big\} \cdot 1_{H(B)}. \nonumber
\end{equation}

\noindent Under the Max-Weight scheduling policy, queue $l$ receives no service until time slot $T_B$. Moreover, queue $h$ is served at unit rate. So, for sample paths in $H(B)$,
\begin{equation}
B-(T_B-1) \leq Q_h(T_B) \leq \sum_{g \neq h} Q_g(T_B) = \sum_{g \neq h} \sum_{t=1}^{T_B-1} A_g(t). \nonumber
\end{equation}

\par A direct consequence of the Strong Law of Large Numbers is the existence of positive constants $\epsilon$ and $\delta$, such that the set of sample paths:
\begin{equation}
\Delta = \Big\{ \Big| \sum_{\tau=1}^{t} A_g(\tau)-\lambda_g \Big| \leq \epsilon \cdot t+\delta, \ \forall t \in N, \ \forall g \neq h \Big\}, \nonumber
\end{equation}
\noindent has positive probability (see Lemma 3 in Appendix 1.2.) We denote by $\tilde{H}(B)$ the set of sample paths $\Delta \cap H(B)$. Due to the IID nature of the arriving traffic, $P(\tilde{H}(B)) = P(\Delta) \cdot P(H(B))$. For sample paths in $\tilde{H}(B)$, we have:
\begin{equation}
T_B-1 \geq \frac{B - (F-1)\cdot\delta}{\sum_{g \neq h}(\lambda_g+\epsilon)+1}. \nonumber
\end{equation}

\noindent Moreover,
\begin{equation}
Q_l(T_B) = \sum_{t=1}^{T_B-1} A_l(t) \geq (\lambda_l-\epsilon) \cdot (T_B-1) - \delta. \nonumber
\end{equation}

\noindent Consequently, for sample paths in $\tilde{H}(B)$ there exist positive constants $c$ and $B_0$, such that:
\begin{equation}
Q_l(T_B) \geq cB, \qquad \forall B \geq B_0. \nonumber
\end{equation}

\par Since at most one packet from queue $l$ can be served at each time slot, the length of queue $l$ is at least $cB/2$ over a time period of length at least $cB/2$ time slots. This implies that the aggregate reward $R^M_{agg}$, i.e., the reward accumulated over a renewal period, satisfies the lower bound
\begin{equation}
R^M_{agg} \cdot 1_{\{B \geq B_0\}} \cdot 1_{\tilde{H}(B)} \geq \min \Big\{ \Big( \frac{cB}{2}\Big)^2 \cdot 1_{\{B \geq B_0\}}, M^2 \Big\} \cdot 1_{\tilde{H}(B)}. \nonumber
\end{equation}

\noindent Then, the expected aggregate reward satisfies
\begin{align}
E[R^M_{agg}] &\geq \sum_{B=1}^{\infty} E[R^M_{agg} \cdot 1_{\{B \geq B_0\}} \cdot 1_{\tilde{H}(B)}] \nonumber \\
&\geq P(\Delta) \cdot \prod_{g \neq h} P(A_g(0)=0) \cdot \sum_{B=1}^{\infty} \min \Big\{ \Big( \frac{cB}{2}\Big)^2 \cdot 1_{\{B \geq B_0\}}, M^2 \Big\} \cdot P(A_h(0)=B). \nonumber
\end{align}

\noindent So, there exists a positive constant $c'$, such that
\begin{equation}
E[R^M_{agg}]\geq c' \cdot E \Big[\min \Big\{ \Big( \frac{c A_h(0)}{2}\Big)^2 \cdot 1_{\{A_h(0) \geq B_0\}}, M^2 \Big\} \Big]. \nonumber
\end{equation}

\noindent Finally, Lemma 4 (see Appendix 1.3) applied to $Y= (1/4)c^2 A^2_h(0) \cdot 1_{\{A_h(0) \geq B_0\}}$, implies that $E[Q_l]=\infty$.


\bigskip\section*{Appendix 4 - Proof of Proposition 1}

\par Consider the single-hop queueing network of Figure 3 under the Max-Weight scheduling policy. Assume that traffic flow 1 is heavy-tailed, traffic flows 2 and 3 are light-tailed, and also that $\lambda_2 > (1+\lambda_1-\lambda_3)/2$. We will show that
\begin{equation}
E[Q_2]=\infty. \nonumber
\end{equation}
\noindent Combined with Lemma 1, this will imply the delay instability of queue 2.

\par Our proof is based on renewal theory, using a strategy similar to the one in the proof of Theorem 2. The time slots that initiate busy periods of the network constitute regeneration epochs. Denote by $X_i$ the length of the $i^{th}$ cycle. The random variables $\{X_i; \ i \in N\}$ can be viewed as IID copies of some nonnegative random variable $X$, with finite first moment; this is because the network is stable under the Max-Weight policy and the empty state is recurrent. 

\par We define an instantaneous reward on this renewal process:
\begin{equation}
R^M(t) = \min\{Q_2(t),M\}, \qquad \ t \in Z_+, \nonumber
\end{equation}
\noindent where $M$ is a positive integer. 

\par Without loss of generality, assume that a renewal period of the system starts at time slot 0. Consider the set of sample paths of the network, where at time slot 0, queue 1 receives a file of size $B$ packets, and all other queues receive no traffic; we denote this set of sample paths by $H(B)$. Clearly, the event $H(B)$ has positive probability, as long as $B$ is in the support of $A_1(0)$, which we henceforth assume:
\begin{equation}
P(H(B))= P(A_1(0)=B) \cdot P(A_2(0)=0) \cdot P(A_3(0)=0). \nonumber
\end{equation}

\par Our \textbf{proof strategy} is as follows: initially, queue 3 does not receive service under Max-Weight, so it starts building up. At the time slot when the service switches from schedule $\{1,2\}$ to schedule $\{3\}$, and if the arrival processes of all traffic flows exhibit their ``average'' behavior, queues 1 and 3 are proportional to $B$, whereas queue 2 remains small. Then, Max-Weight will start draining the weights of the two feasible schedules at roughly the same rate, until one of them empties. Let $\mu_f$ denote the departure rate from queue $f$ during this period. Roughly speaking, the departure rates are the solution to the following system of linear equations:
\begin{align}
\lambda_1 + \lambda_2 - \mu_1 - \mu_2 &= \lambda_3 - \mu_3 \nonumber \\
\mu_1 + \mu_3 &= 1 \nonumber \\
\mu_1 &= \mu_2. \nonumber
\end{align}

\noindent The last two equations follow from the facts that Max-Weight is a work-conserving policy, and that queues 1 and 2 are served simultaneously. If the rate at which traffic arrives to queue 2 is greater than the rate at which it departs from it, i.e.,
\begin{equation}
\lambda_2 > \mu_2 = \frac{1+\lambda_1+\lambda_2-\lambda_3}{3}, \nonumber
\end{equation}

\noindent or, equivalently,
\begin{equation}
\lambda_2 > \frac{1+\lambda_1-\lambda_3}{2}, \nonumber
\end{equation}

\noindent then queue 2 builds up during this time period, which is proportional to $B$. This implies that $E[Q_2]=\infty$, since $B$ is heavy-tailed distributed.

\par Throughout the proof we use the following shorthand notation: we say that a random variable $X$ scales at least linearly with $B$ on the event $H$, and write $X=\Omega_H(B)$, if there exist positive constants $k$ and $k'$ (possibly depending on the event $H$), such that $X \geq k \cdot B - k'$, for all sample paths in $H$.

\par We break the proof into four steps.

\medskip\underline{Step 1}

\par For sample paths in $H(B)$, denote by $T^1_B$ the first time slot, starting from 0, when the length of queue 3 becomes greater than or equal to the sum of the lengths of queues 1 and 2:
\begin{equation}
T^1_B = \min \{t>0 \mid Q_3(t) \geq Q_1(t)+Q_2(t) \} \cdot 1_{H(B)}. \nonumber
\end{equation}

\noindent The first part of the proof is to show that $Q_1(T^1_B)$ and $Q_3(T^1_B)$ scale at least linearly with $B$, provided all arrival processes exhibit their ``average'' behavior.

\par Under the Max-Weight scheduling policy, queue 3 receives no service until time-slot $T^1_B$. Moreover, the server of the system has unit service rate. So, for sample paths in $H(B)$:
\begin{equation}
Q_1(T^1_B) \leq Q_1(T^1_B) + Q_2(T^1_B) \leq Q_3(T^1_B). \nonumber
\end{equation}

\noindent A direct consequence of the Strong Law of Large Numbers is the existence of positive constants $\delta$ and $\epsilon$, such that the set of sample paths:
\begin{equation}
\Delta(B) = \Big\{ (\lambda_f-\epsilon)t-\delta \leq \sum_{\tau=1}^{t} A_f(\tau) \leq (\lambda_f+\epsilon)t+\delta, \ \forall t \in \{1,\ldots,T_B^1-1\}, \ \forall f \in \{1,2,3\} \Big\}, \nonumber
\end{equation}

\noindent has probability bounded away from 0, uniformly over all $B$ (see  Lemma 3 in Appendix 1.2.) Note that $\epsilon$ can be chosen arbitrarily small. Similarly,
\begin{equation}
\Delta = \Big\{ (\lambda_f-\epsilon)t-\delta \leq \sum_{\tau=1}^{t} A_f(\tau) \leq (\lambda_f+\epsilon)t+\delta, \ \forall t \in N, \ \forall f \in \{1,2,3\} \Big\}, \nonumber
\end{equation}

\noindent has also probability bounded away from 0. Denote by $\tilde{H}(B)$ the set of sample paths $H(B) \cap \Delta(B)$, and observe that $H(B) \cap \Delta(B) \supset H(B) \cap \Delta$. Then, the IID nature of the arriving traffic implies:
\begin{equation}
P(\tilde{H}(B))=P(H(B) \cap \Delta(B)) \geq P(H(B) \cap \Delta) =  P(H(B)) \cdot P(\Delta)>0. \nonumber
\end{equation}

\par For sample paths in $\tilde{H}(B)$, we have:
\begin{equation}
Q_1(T^1_B) \geq B -(T^1_B-1) + (\lambda_1-\epsilon) \cdot (T^1_B-1) -\delta. \nonumber
\end{equation}

\noindent Moreover,
\begin{equation}
Q_3(T^1_B) = \sum_{t=1}^{T^1_B-1} A_3(t) \leq (\lambda_3+\epsilon) \cdot (T^1_B-1) + \delta. \nonumber
\end{equation}

\noindent Consequently, since $Q_1(T^1_B) \leq Q_3(T^1_B)$, we obtain:
\begin{equation}
T^1_B - 1 \geq \frac{B - 2 \delta}{1+\lambda_3-\lambda_1 + 2 \epsilon}. \tag{4}
\end{equation}

\noindent Therefore,
\begin{align}
Q_3(T^1_B) = \sum_{t=1}^{T^1_B-1} A_3(t) &\geq (\lambda_3-\epsilon) \cdot (T^1_B-1) - \delta, \nonumber \\
&\geq (\lambda_3-\epsilon) \cdot \frac{B - 2 \delta}{1+\lambda_3-\lambda_1 + 2 \epsilon} - \delta, \nonumber
\end{align}

\noindent which implies that $Q_3(T^1_B)=\Omega_{\tilde{H}(B)}(B)$.

\par Coming to queue 2, it can be verified that for sample paths in $\tilde{H}(B)$ and for any subinterval $\{\tau_0,\ldots,\tau_1\}$ of $\{1,\ldots,T^1_B\}$: 
\begin{equation}
\sum_{t=\tau_0}^{\tau_1-1} A_2(t) \leq (\lambda_2 + 2 \epsilon) \cdot (\tau_1-\tau_0) + 2 \delta. \nonumber
\end{equation} 

\noindent If $\epsilon$ is chosen sufficiently small, such that $\lambda_2 + 2 \epsilon<1$, then

\begin{equation}
Q_2(T^1_B) \leq A_2(T^1_B-1) + 2 \delta \leq \lambda_2 + 2 \epsilon (T^1_B-1) + 4 \delta, \tag{5}
\end{equation}

\noindent since queue 2 gets served whenever it is nonempty throughout the period $\{1,\ldots,T^1_B-1\}$. This shows that, essentially, $Q_2(T^1_B)$ does not scale with $B$.

\par We finally develop a lower bound on $Q_1(T^1_B)$. By definition,
\begin{equation}
Q_3(T^1_B-1) < Q_1(T^1_B-1)+Q_2(T^1_B-1). \tag{6}
\end{equation}

\noindent By arguing similarly to Eq. (5), it can be verified that
\begin{equation}
Q_2(T^1_B-1) \leq \lambda_2 + 2 \epsilon (T^1_B-2) + 4 \delta. \tag{7}
\end{equation}

\noindent Eq. (6) and (7), combined with the fact that queue 1 is served at each time slot until $T^1_B$, imply:
\begin{equation}
Q_3(T^1_B-1) \leq B -(T^1_B-2) + (\lambda_1+3 \epsilon) \cdot (T^1_B-2) + 5 \delta_2 + \lambda_2. \tag{8}
\end{equation}

\noindent Moreover, for sample paths in $\tilde{H}(B)$
\begin{equation}
Q_3(T^1_B-1) \geq (\lambda_3 -\epsilon) \cdot (T^1_B-2) -\delta. \tag{9}
\end{equation}

\noindent Eq. (8) and (9) give:
\begin{equation}
T^1_B-1 < \frac{B + 6 \delta + \lambda_2}{1+\lambda_3-\lambda_1- 4 \epsilon}+ 1, \nonumber
\end{equation}

\noindent which, combined with Eq. (4), results in:
\begin{align}
Q_1(T^1_B) &\geq B - (T^1_B-1) + (\lambda_1-\epsilon) \cdot (T^1_B-1) - \delta \nonumber \\
&> B - \frac{B + 6 \delta + \lambda_2}{1+\lambda_3-\lambda_1- 4 \epsilon} - 1 \nonumber \\
&+ (\lambda_1-\epsilon) \cdot \frac{B- 2 \delta}{1+\lambda_3-\lambda_1+ 2 \epsilon} -\delta. \nonumber
\end{align}

\noindent It follows that $Q_1(T^1_B)=\Omega_{\tilde{H}(B)}(B)$, provided $\epsilon$ is chosen sufficiently small.

\par To summarize: at time slot $T^1_B$, queues 1 and 3 are proportional to $B$, while queue 2 has remained small.

\medskip\underline{Step 2}

\par Now denote by $T^2_B$ the first time slot after $T_B^1$, that either queue 1 or queue 3 becomes empty:
\begin{equation}
T^2_B = \min \{t>T_B^1 \mid Q_1(t) \cdot Q_3(t)=0 \} \cdot 1_{\tilde{H}(B)}. \nonumber
\end{equation}

\noindent The second part of the proof is to show that if the arrival processes exhibit their ``average'' behavior, then at time slot $T^2_B$, the length of queue 3 is, roughly speaking, no larger than the sum of the lengths of queues 1 and 2. 

\par For the same constants $\delta$ and $\epsilon$ defined in Step 1, the set of sample paths:
\begin{equation}
\Delta'(B) = \Big\{ (\lambda_f-\epsilon)t-\delta \leq \sum_{\tau=T_B^1}^{t} A_f(\tau) \leq (\lambda_f+\epsilon)t+\delta, \ \forall t \in \{T_B^1,\ldots,T_B^2-1\}, \ \forall f \in \{1,2,3\} \Big\}, \nonumber
\end{equation}

\noindent has probability bounded away from 0. We denote by $\hat{H}(B)$ the set of sample paths $\tilde{H}(B) \cap \Delta'(B)$. Due to the IID nature of the arriving traffic:
\begin{equation}
P(\hat{H}(B)) \geq P(H(B)) \cdot P(\Delta)^2>0. \nonumber
\end{equation}

\par We will show that for sample paths in $\hat{H}(B)$:
\begin{equation}
Q_3(T^2_B) \leq Q_1(T^2_B) + Q_2(T^2_B) + 2 \epsilon (T^2_B-T^1_B) + 2 \delta + 3. \tag{10}
\end{equation}

\par First, notice that queues 1 and 3 cannot empty at the same time slot, since they cannot be served simultaneously. Therefore, we have two possible cases: if $Q_3(T^2_B)=0$, then Eq. (10) is trivially satisfied. Otherwise, suppose that $Q_1(T^2_B)=0$. Then, $S_1(T^2_B-1)=S_2(T^2_B-1)=1$, while $S_3(T^2_B-1)=0$. For sample paths in $\hat{H}(B)$ we have:
\begin{align}
Q_3(T^2_B) &= Q_3(T^2_B-1) + A_3(T^2_B-1) \nonumber \\
&\leq Q_3(T^2_B-1) + \lambda_3 + 2 \epsilon \cdot (T^2_B-T^1_B) + 2 \delta. \tag{11}
\end{align}

\noindent Moreover, under the Max-Weight policy:
\begin{equation}
Q_3(T^2_B-1) \leq Q_1(T^2_B-1) + Q_2(T^2_B-1). \tag{12}
\end{equation}

\noindent Finally,
\begin{equation}
Q_1(T^2_B-1) + Q_2(T^2_B-1) - 2 \leq Q_1(T^2_B) + Q_2(T^2_B), \nonumber
\end{equation}

\noindent which, in turn, gives:
\begin{equation}
Q_1(T^2_B-1) + Q_2(T^2_B-1) + 2 \epsilon \cdot (T^2_B-T^1_B) + 2 \delta + 1 \leq Q_1(T^2_B) + Q_2(T^2_B) + 2 \epsilon \cdot (T^2_B-T^1_B) + 2 \delta + 3. \tag{13}
\end{equation}

\noindent Since $\lambda_3<1$, Eq. (11)-(13) imply that Eq. (10) holds.

\medskip\underline{Step 3}

\par The third part of the proof uses the results of Steps 1 and 2 in order to show that, for the sample paths of interest and if $\lambda_2>(1+\lambda_1-\lambda_3)/2$, then $Q_2(T^2_B)=\Omega_{\hat{H}(B)}(B)$.

\par By definition,
\begin{equation}
Q_3(T^1_B) \geq Q_1(T^1_B) + Q_2(T^1_B). \nonumber
\end{equation}

\noindent By substituting the two sides of Eq. (10), we get:
\begin{equation}
Q_3(T^2_B)-Q_3(T^1_B) \leq Q_1(T^2_B)-Q_1(T^1_B) + Q_2(T^2_B)-Q_2(T^1_B) + 2 \epsilon (T^2_B-T^1_B) + 2 \delta+3. \tag{14}
\end{equation}

\par For sample paths in $\hat{H}(B)$ define the random variables:
\begin{equation}
\mu_f = \Big( \frac{1}{T^2_B-T^1_B} \cdot \sum_{t=T^1_B}^{T^2_B-1} S_f(t) \Big) \cdot 1_{\hat{H}(B)}, \qquad f \in \{1,2,3\}, \nonumber
\end{equation}

\noindent which are the average service rates to each queue during the interval $\{T^1_B,\ldots,T^2_B-1\}$. Notice that
\begin{equation}
\mu_1=\mu_2, \tag{15}
\end{equation}

\noindent and also
\begin{equation}
\mu_1+\mu_3=1. \tag{16}
\end{equation}

\par Since both queues 1 and 3 are nonempty during the inerval $\{T^1_B,\ldots,T^2_B-1\}$, we have:
\begin{align}
Q_1(T^2_B)-Q_1(T^1_B) &\leq (\lambda_1+\epsilon-\mu_1) \cdot (T^2_B-T^1_B)+\delta, \tag{17} \\
Q_3(T^2_B)-Q_3(T^1_B) &\geq (\lambda_3-\epsilon-\mu_3) \cdot (T^2_B-T^1_B)-\delta. \tag{18}
\end{align}

\par Eqs. (14), (17), and (18) imply:
\begin{equation}
(\lambda_3-\epsilon-\mu_3) \cdot (T^2_B-T^1_B)-\delta \leq (\lambda_1+\epsilon-\mu_1) \cdot (T^2_B-T^1_B)+\delta + Q_2(T^2_B)-Q_2(T^1_B) + 2 \epsilon (T^2_B-T^1_B) + 2 \delta+3, \nonumber
\end{equation}

\noindent Using Eq. (16) and collecting terms:
\begin{align}
-\mu_1 \cdot (T^2_B-T^1_B) &\geq - \Big( \frac{1+\lambda_1-\lambda_3+4 \epsilon}{2} \Big) \cdot (T^2_B-T^1_B) + \frac{Q_2(T^1_B)-Q_2(T^2_B)}{2} - \frac{4 \delta +3}{2} \nonumber \\
&\geq - \Big( \frac{1+\lambda_1-\lambda_3+ 4 \epsilon}{2} \Big) \cdot (T^2_B-T^1_B) - \frac{Q_2(T^2_B)}{2} - \frac{4 \delta+3}{2}. \nonumber
\end{align}

\noindent Then, for sample paths in $\hat{H}(B)$, the queue length $Q_2(T^2_B)$ is bounded from below by:
\begin{align}
Q_2(T^2_B) &\geq (\lambda_2 - \epsilon -\mu_1) \cdot (T^2_B-T^1_B) -\delta \nonumber \\
&\geq \Big( \lambda_2 - \frac{1+\lambda_1-\lambda_3}{2} - 3 \epsilon \Big) \cdot (T^2_B-T^1_B) -\frac{Q_2(T^2_B)}{2}-\frac{6 \delta+3}{2}. \nonumber
\end{align}

\noindent In the first inequality we have also used Eq. (15). Therefore,
\begin{equation}
Q_2(T^2_B) \geq \frac{2}{3} \cdot \Big( \lambda_2 - \frac{1+\lambda_1-\lambda_3}{2} - 3 \epsilon \Big) \cdot (T^2_B-T^1_B) -2 \delta - 1. \nonumber
\end{equation}

\noindent If $\lambda_2>(1+\lambda_1-\lambda_3)/2$, the constant $\epsilon$ can be chosen sufficiently small, so that:
\begin{equation}
\lambda_2 - \frac{1+\lambda_1-\lambda_3}{2} - 3 \epsilon>0. \nonumber
\end{equation}

\par A final observation is that the duration of the interval $\{T^1_B,\ldots,T^2_B-1\}$ is bounded from below by $\min\{Q_1(T^1_B),Q_3(T^1_B)\}$, because both queues are served at unit rate. So,
\begin{equation}
T^2_B-T^1_B=\Omega_{\tilde{H}(B)}(B). \nonumber
\end{equation}

\noindent Consequently,
\begin{equation}
Q_2(T^2_B)=\Omega_{\hat{H}(B)}(B). \tag{19}
\end{equation}

\medskip\underline{Step 4}

\par In Step 3 we showed that for sample paths in $\hat{H}(B)$, queue 2 builds up to the order of $B$. In the fourth and final step of the proof, we show that this implies that the expected steady-state length of queue 2 is infinite.

\par Eq. (19) implies that for sample paths in $\hat{H}(B)$, there exist positive constants $c$ and $B_0$, such that:
\begin{equation}
Q_2(T^2_B) \geq cB, \qquad \forall B \geq B_0. \nonumber
\end{equation}

\par Since at most one packet from queue 2 can be served at each time slot, the length of queue 2 is at least $cB/2$ packets over a time period of length at least $cB/2$ time slots. Hence, the aggregate reward $R^M_{agg}$, i.e., the reward accumulated over a renewal period, satisfies the lower bound
\begin{equation}
R^M_{agg} \cdot 1_{\{B \geq B_0\}} \cdot 1_{\hat{H}(B)} \geq \min \Big\{ \Big( \frac{cB}{2}\Big)^2 \cdot 1_{\{B \geq B_0\}}, M^2 \Big\} \cdot 1_{\hat{H}(B)}. \nonumber
\end{equation}

\noindent Then, the expected aggregate reward is bounded by
\begin{align}
E[R^M_{agg}] &\geq \sum_{B=1}^{\infty} E[R^M_{agg} \cdot 1_{\{B \geq B_0\}} \cdot 1_{\hat{H}(B)}] \nonumber \\
&\geq P(\Delta)^2 \cdot P(A_2(0)=0) \cdot P(A_3(0)=0) \nonumber \\
&\cdot \sum_{B=1}^{\infty} \min \Big\{ \Big( \frac{cB}{2}\Big)^2 \cdot 1_{\{B \geq B_0\}}, M^2 \Big\} \cdot P(A_1(0)=B). \nonumber
\end{align}

\noindent So, there exists a positive constant $c'$, such that
\begin{equation}
E[R^M_{agg}] \geq c' \cdot E \Big[ \min \Big\{ \Big( \frac{c A_1(0)}{2}\Big)^2 \cdot 1_{\{A_1(0) \geq B_0\}}, M^2 \Big\} \Big]. \nonumber
\end{equation}

\noindent Finally, Lemma 4 (see Appendix 1.3) applied to $Y= (1/4)c^2 A^2_1(0) \cdot 1_{\{A_1(0) \geq B_0\}}$, implies that $E[Q_2]=\infty$.


\bigskip\section*{Appendix 5 - Proof of Theorem 3}

\par Consider a set of feasible schedules $\{\sigma^k;\ k=1,\ldots,k^*\}$ such that:
\begin{equation}
\bigcup_{k=1}^{k^*} \sigma^k = \{1,\ldots,F\}. \nonumber
\end{equation}

\noindent (The admissibility of the arriving traffic implies that such a set of feasible schedules exists.) 

\par By the definition of the intensity parameter $\rho \in (0,1)$, there exist nonnegative numbers $\zeta_i, \ i=1,\ldots,I$, adding up to 1, and feasible schedules $\tilde{s}^i, \ i=1,\ldots,I$, such that:
\begin{equation}
\lambda \leq \rho \cdot \sum_{i=1}^I \zeta_i \cdot \tilde{s}^i. \nonumber
\end{equation}

\par Notice that
\begin{equation}
\Big( (1-\rho) \cdot \sum_{k=1}^{k^*} \frac{1}{k^*} \cdot \sigma^k + \rho \cdot \sum_{i=1}^I \zeta_i \cdot \tilde{s}^i \Big) \in \overline{\Lambda}, \nonumber
\end{equation}
\noindent where $\overline{\Lambda}$ denotes the closure of the set $\Lambda$. This is because we have a convex combination of $(I+k^*)$ feasible schedules, and the stability region is known to be a convex set; see Section 3.2 of \cite{GNT06}. 
Moreover,
\begin{align}
(1-\rho) \cdot \sum_{k=1}^{k^*} \frac{1}{k^*} \cdot \sigma^k &= \frac{1-\rho}{k^*} \cdot \sum_{k=1}^{k^*} \sigma^k \nonumber \\
&\geq \frac{1-\rho}{k^*} \cdot 1_F, \nonumber
\end{align}
\noindent where $1_F$ denotes the $F$-dimensional vector of ones. 

\par A well-known monotonicity property of the stability region is the following: if $0 \leq \lambda' \leq \lambda''$ componentwise, and $\lambda'' \in \Lambda$, then $\lambda' \in \Lambda$. Using this property, we have:
\begin{equation}
\Big( \frac{1-\rho}{k^*} \cdot 1_F + \lambda \Big) \in \overline{\Lambda}. \nonumber
\end{equation}
This, in turn, implies the existence of nonnegative numbers $\theta_j, \ j=1,\cdots,J$, adding up to 1, and of feasible schedules $s^j=(s_f^j), \ j=1,\cdots,J$, such that:
\begin{equation}
\lambda_f \leq \sum_{j=1}^J \theta_j \cdot s_f^j - \frac{1-\rho}{k^*}, \qquad \forall f \in \{1,\ldots,F\}. \tag{20}
\end{equation}

\medskip\par Under the Max-Weight-$\alpha$ scheduling policy the sequence $\{Q(t);\ t \in Z_+\}$ is a time-homogeneous, irreducible, and aperiodic Markov chain on the countable state-space $Z_+^F$. We will prove that this Markov chain is also positive recurrent, and we will establish upper bounds for the $\alpha$-moments of the steady-state queue lengths, provided that $E[A_f^{\alpha_f+1}(0)]<\infty$, for all $f \in \{1,\ldots,F\}$.

\par Consider the Lyapunov function
\begin{equation}
V(Q) = \sum_{f=1}^F \frac{1}{\alpha_f+1} Q_f^{\alpha_f+1}. \nonumber
\end{equation}
We have 
\begin{equation}
E[V(Q(t+1)) \mid Q(t)] = \sum_{f=1}^F E\Big[\frac{1}{\alpha_f+1} (Q_f(t)+\Delta_f(t))^{\alpha_f+1} \ \Big| \ Q(t)\Big], \nonumber
\end{equation}
where
\begin{equation}
\Delta_f(t) = A_f(t) - S_f(t) \cdot 1_{\{Q_f(t)>0\}}. \nonumber
\end{equation}

\noindent Throughout the proof we use the shorthand notation 
\begin{equation}
V_f(Q_f(t)) = \frac{1}{\alpha_f+1} Q_f^{\alpha_f+1}(t). \nonumber
\end{equation}
We consider the conditional expectation of the terms $V_f(Q_f(t +1))$, distinguishing between two cases.

\medskip\par i) $\alpha_f \leq 1$: Consider the zeroth order Taylor expansion around $Q_f(t)$ (i.e., the mean value theorem):
\begin{equation}
\frac{1}{\alpha_f+1} (Q_f(t)+\Delta_f(t))^{\alpha_f+1} = \frac{1}{\alpha_f+1} Q_f^{\alpha_f+1}(t) + \Delta_f(t)\cdot \xi(t)^{\alpha_f}, \nonumber
\end{equation}

\noindent for some $\xi(t) \in\ [Q_f(t)-S_f(t) \cdot 1_{\{Q_f(t)>0\}},Q_f(t)+A_f(t)]$. Thus,
\begin{align}
V_f(Q_f(t+1)) = V_f(Q_f(t)) + \Delta_f(t)\cdot \xi(t)^{\alpha_f}, \nonumber
\end{align}

\noindent and
\begin{equation}
E[V_f(Q_f(t+1)) \mid Q(t)] = V_f(Q_f(t)) + E[\Delta_f(t)\cdot \xi(t)^{\alpha_f} \mid Q(t)].  \nonumber
\end{equation}

\par Consider the event $\Gamma_f(t)=\{\Delta_f(t) < 0\}$ and its complement. We have:
\begin{align}
E[V_f(Q_f(t+1)) \mid Q(t)] &\leq V_f(Q_f(t)) + E[\Delta_f(t) \cdot (Q_f(t)+A_f(t))^{\alpha_f} \cdot 1_{\{\Gamma_f^c(t)\}} \mid Q(t)] \nonumber \\
&+ E[\Delta_f(t) \cdot (Q_f(t)-S_f(t) \cdot 1_{\{Q_f(t)>0\}})^{\alpha_f} \cdot 1_{\{\Gamma_f(t)\}} \mid Q(t)]. \tag{21}
\end{align}

\noindent Since $Q_f(t),\ Q_f(t)-S_f(t) \cdot 1_{\{Q_f(t)>0\}}$, and $A_f(t)$ are nonnegative numbers and $\alpha_f \in (0,1]$, it can be verified that 
\begin{equation}
(Q_f(t)+A_f(t))^{\alpha_f} \leq Q_f^{\alpha_f}(t) + A_f^{\alpha_f}(t). \tag{22}
\end{equation}

\noindent Moreover, since they are also integers,
\begin{equation}
(Q_f(t)-S_f(t) \cdot 1_{\{Q_f(t)>0\}})^{\alpha_f} \geq Q_f^{\alpha_f}(t) - S_f(t) \cdot 1_{\{Q_f(t)>0\}}. \tag{23}
\end{equation}

\noindent Eqs. (21)-(23) imply that
\begin{align}
E[V_f(Q_f(t+1)) \mid Q(t)] &\leq V_f(Q_f(t)) + E[\Delta_f(t) \mid Q(t)] \cdot Q_f^{\alpha_f}(t) \nonumber \\
&+ E[\Delta_f(t) \cdot A_f^{\alpha_f}(t) \cdot 1_{\{\Gamma_f^c(t)\}} \mid Q(t)] \nonumber \\
&+ E[-\Delta_f(t) \cdot S_f(t) \cdot 1_{\{Q_f(t)>0\}} \cdot 1_{\{\Gamma_f(t)\}}\mid Q(t)]. \nonumber
\end{align}

\noindent If $\Delta_f(t)<0$, which is denoted by the event $\Gamma_f(t)$, then $-\Delta_f(t) \leq 1$. Also, if $\Delta_f(t) \geq 0$, which is denoted by the event $\Gamma_f^c(t)$, then $\Delta_f(t) \leq A_f(t)$, so that $\Delta_f(t) \cdot A_f^{\alpha_f}(t) \leq A_f^{\alpha_f+1}(t)$. Consequently,
\begin{equation}
E[V_f(Q_f(t+1)) \mid Q(t)] \leq V_f(Q_f(t)) + E[\Delta_f(t) \mid Q(t)] \cdot Q_f^{\alpha_f}(t) + E[A_f^{\alpha_f+1}(t) \cdot 1_{\{\Gamma_f^c(t)\}}\mid Q(t)] + 1. \nonumber
\end{equation}

\noindent Finally, the fact that the random variables $\{A_f(t);\ t \in Z_+\}$ are IID gives:
\begin{equation}
E[V_f(Q_f(t+1)) \mid Q(t)] \ \leq \  V_f(Q_f(t)) + E[\Delta_f(t) \mid Q(t)] \cdot Q_f^{\alpha_f}(t) + E[A_f^{\alpha_f+1}(0)] + 1. \nonumber
\end{equation}

\noindent The inequality above implies that
\begin{equation}
E[V_f(Q_f(t+1)) \mid Q(t)] \leq V_f(Q_f(t)) + E[\Delta_f(t) \mid Q(t)] \cdot Q_f^{\alpha_f}(t) + \frac{1-\rho}{2 k^*} \cdot Q_f^{\alpha_f}(t) + E[A_f^{\alpha_f+1}(0)] + 1. \tag{24}
\end{equation}

\bigskip\par ii) $\alpha_f > 1$: Consider the first order Taylor expansion around $Q_f(t)$:
\begin{equation}
\frac{1}{\alpha_f+1} (Q_f(t)+\Delta_f(t))^{\alpha_f+1} = \frac{1}{\alpha_f+1} Q_f(t)^{\alpha_f+1} + \Delta_f(t) \cdot Q_f^{\alpha_f}(t) + \frac{\Delta_f^2(t)}{2} \cdot \alpha_f \cdot \xi(t)^{\alpha_f-1}, \nonumber
\end{equation}

\noindent for some $\xi(t) \in [Q_f(t)-S_f(t) \cdot 1_{\{Q_f(t)>0\}},Q_f(t)+A_f(t)]$. Then,
\begin{equation}
E[V_f(Q_f(t+1)) \mid Q(t)] \ =  V_f(Q_f(t)) + E[\Delta_f(t) \mid Q(t)] \cdot Q_f^{\alpha_f}(t) + E\Big[\frac{\Delta_f^2(t)}{2} \cdot \alpha_f \cdot \xi(t)^{\alpha_f-1} \ \Big| \ Q(t)\Big]. \tag{25}
\end{equation}

\par Since $\Delta_f^2(t) \cdot \alpha_f \geq 0$ and $\alpha_f-1 \geq 0$, the last term can be bounded from above by
\begin{equation}
E\Big[\frac{\Delta_f^2(t)}{2} \cdot \alpha_f \cdot \xi(t)^{\alpha_f-1} \ \Big| \ Q(t)\Big] \leq E\Big[\frac{\Delta_f^2(t)}{2} \cdot \alpha_f \cdot (Q_f(t)+A_f(t))^{\alpha_f-1} \ \Big| \ Q(t)\Big]. \tag{26}
\end{equation}

\noindent Moreover, it is easy to verify that for $\alpha_f \geq 1$,
\begin{equation}
(Q_f(t)+A_f(t))^{\alpha_f-1} \leq 2^{\alpha_f-1} \cdot(Q_f^{\alpha_f-1}(t) + A_f^{\alpha_f-1}(t)), \tag{27}
\end{equation}

\noindent and also that
\begin{equation}
\Delta_f^2(t) \leq A_f^2(t) + 1. \tag{28}
\end{equation}

\noindent Eqs. (26)-(28) imply that
\begin{align}
E\Big[\frac{\Delta_f^2(t)}{2} \cdot \alpha_f \cdot \xi^{\alpha_f-1} \ \Big| \ Q(t)\Big] &\leq 2^{\alpha_f-2} \cdot \alpha_f \cdot \Big( E[A_f^2(t)] + 1 \Big) \cdot Q_f^{\alpha_f-1}(t) \nonumber \\
&+ 2^{\alpha_f-2} \cdot \alpha_f \cdot \Big( E[A_f^{\alpha_f+1}(t)] + E[A_f^{\alpha_f-1}(t)] \Big) \nonumber \\
&\leq K \cdot Q_f^{\alpha_f-1}(t)+K, \tag{29}
\end{align}

\noindent where $K=2^{\alpha_f-1} \cdot \alpha_f \cdot \Big( E[A_f^{\alpha_f+1}(0)]+1 \Big)$. Then, Eqs. (25) and (29) imply that
\begin{align}
E[V_f(Q_f(t+1)) \mid Q(t)] &\leq  V_f(Q_f(t)) + E[\Delta_f(t) \mid Q(t)] \cdot Q_f^{\alpha_f}(t) + K \cdot Q_f^{\alpha_f-1}(t) + K \nonumber \\
&=  V_f(Q_f(t)) + E[\Delta_f(t) \mid Q(t)] \cdot Q_f^{\alpha_f}(t) + \frac{1-\rho}{2 k^*} \cdot Q_f^{\alpha_f}(t) \nonumber \\
&+ \Big(K \cdot Q_f^{\alpha_f-1}(t) - \frac{1-\rho}{2 k^*} \cdot Q_f^{\alpha_f}(t) + K \Big). \tag{30}
\end{align}

\par Our goal is to bound from above the last term of the right-hand side of Eq. (30). Relaxing the constraint that $Q_f(t)$ has to be an integer, we have:
\begin{equation}
K \cdot Q_f^{\alpha_f-1}(t) - \frac{1-\rho}{2 k^*} \cdot Q_f^{\alpha_f}(t) + K \leq \max_{x \in \Re_+} \Big\{ K \cdot x^{\alpha_f-1} - \frac{1-\rho}{2k^*} \cdot x^{\alpha_f} + K \Big\}, \qquad \forall t \in Z_+. \tag{31}
\end{equation}

\noindent It can be verified that the optimization problem in the right-hand side has the unique solution $x^*=\frac{2 k^* K}{1-\rho} \cdot \frac{\alpha_f-1}{\alpha_f}$. Therefore, the optimal value is
\begin{equation}
K^{\alpha_f} \cdot \Big( \frac{2 k^*}{1-\rho} \Big)^{\alpha_f-1} \cdot \frac{(\alpha_f-1)^{\alpha_f-1}}{\alpha_f^{\alpha_f}} + K \leq K^{\alpha_f} \cdot \Big( \frac{2 k^*}{1-\rho} \Big)^{\alpha_f-1} + K. \tag{32}
\end{equation}

\noindent Eqs. (31) and (32) imply that
\begin{equation}
K \cdot Q_f^{\alpha_f-1}(t) - \frac{1-\rho}{2 k^*} \cdot Q_f^{\alpha_f}(t) + K \leq K^{\alpha_f} \cdot \Big( \frac{2 k^*}{1-\rho} \Big)^{\alpha_f-1} + K, \qquad \forall t \in Z_+. \tag{33}
\end{equation}

\noindent Then, Eqs. (30) and (33) give:
\begin{equation}
E[V_f(Q_f(t+1)) \mid Q(t)] \ =  V_f(Q_f(t)) + E[\Delta_f(t) \mid Q(t)] \cdot Q_f^{\alpha_f}(t) + \frac{1-\rho}{2 k^*} \cdot Q_f^{\alpha_f}(t) + K^{\alpha_f} \cdot \Big( \frac{2 k^*}{1-\rho} \Big)^{\alpha_f-1} + K. \tag{34}
\end{equation}

\bigskip\par Summarizing our findings from cases (i) and (ii), Eqs. (24) and (34) imply that
\begin{align}
E[V_f(Q_f(t+1)) \mid Q(t)] &\leq  V_f(Q_f(t)) + E[\Delta_f(t) \mid Q(t)] \cdot Q_f^{\alpha_f}(t) \nonumber \\
&+ \frac{1-\rho}{2 k^*} \cdot Q_f^{\alpha_f}(t) + H \Big( \rho,k^*,\alpha_f,E[A_f^{\alpha_f+1}(0)] \Big), \nonumber
\end{align}

\noindent for all $f \in \{1,\ldots,F\}$, where
\begin{equation}
H \Big( \rho,k^*,\alpha_f,E[A_f^{\alpha_f+1}(0)] \Big) = \left\{ \begin{array}{ll}
E[A_f^{\alpha_f+1}(0)]+1, & \alpha_f \leq 1, \\
K^{\alpha_f} \cdot \Big( \frac{2 k^*}{1-\rho} \Big)^{\alpha_f-1} + K, & \alpha_f > 1,
\end{array} \right. \nonumber
\end{equation}

\noindent and $K=2^{\alpha_f-1} \cdot \alpha_f \cdot \Big( E[A_f^{\alpha_f+1}(0)]+1 \Big)$. Summing over all $f \in \{1,\ldots,F\}$, gives:
\begin{align}
E[V(Q(t+1)) \mid Q(t)] &\leq V(Q(t)) + \sum_{f=1}^F (\lambda_f - S_f(t) \cdot 1_{\{Q_f(t)>0\}}) \cdot Q_f^{\alpha_f}(t) \nonumber \\
&+ \frac{1-\rho}{2 k^*} \cdot \sum_{f=1}^F Q_f^{\alpha_f}(t) + \sum_{f=1}^F H \Big( \rho,k^*,\alpha_f,E[A_f^{\alpha_f+1}(0)] \Big). \nonumber
\end{align}

\noindent Taking into account Eq. (20), we have:
\begin{align}
E[V(Q(t+1)) \mid Q(t)] &\leq V(Q(t)) -\frac{1-\rho}{2 k^*} \cdot \sum_{f=1}^F Q_f^{\alpha_f}(t) + \sum_{f=1}^F H \Big( \rho,k^*,\alpha_f,E[A_f^{\alpha_f+1}(0)] \Big) \nonumber \\
&+ \sum_{f=1}^F \Big(\sum_{j=1}^{J} \theta_j \cdot s_f^j - S_f(t) \Big) \cdot Q_f^{\alpha_f}(t). \nonumber
\end{align}

\noindent By the definition of the Max-Weight-$\alpha$ scheduling policy, the last term is nonpositive. So,
\begin{equation}
E[V(Q(t+1)) \mid Q(t)] \ \leq \  V(Q(t)) -\frac{1-\rho}{2 k^*} \cdot \sum_{f=1}^F Q_f^{\alpha_f}(t) + \sum_{f=1}^F H \Big( \rho,k^*,\alpha_f,E[A_f^{\alpha_f+1}(0)] \Big). \nonumber
\end{equation}

\noindent Then, the Foster-Lyapunov stability criterion and moment bound (e.g., see Corollary 2.1.5 of \cite{H06}) implies that the sequence $\{Q(t);\ t \in Z_+\}$ converges in distribution. Moreover, its limiting distribution $(Q_f;\ f=1,\ldots,F)$ does not depend on $Q(0)$, and satisfies
\begin{equation}
\sum_{f=1}^F E[Q_f^{\alpha_f}] \leq \frac{2 k^*}{1-\rho} \cdot \sum_{f=1}^F H \Big( \rho,k^*,\alpha_f,E[A_f^{\alpha_f+1}(0)] \Big). \nonumber
\end{equation}
Based on this, it can be verified that the sequence $\{D(k);\ k \in N\}$ is a (possibly delayed) aperiodic and positive recurrent regenerative process. Hence, it also converges in distribution, and its limiting distribution does not depend on $Q(0)$; see \cite{SW93}.


\medskip\begin{thebibliography}{99}

	\bibitem{AKRSVW04}
		M. Andrews, K. Kumaran, K. Ramanan, A. Stolyar, R. Vijayakumar, P. Whiting (2004). Scheduling in a queueing system with asynchronously varying service rates. Probability in the Engineering and Informational Sciences, 18, 191-217.
	  
	\bibitem{BBNZ03}
		S. Borst, O. Boxma, R. Nunez-Queija, B. Zwart (2003). The impact of the service discipline on delay asymptotics. Performance Evaluation, 54, 175-206.
	  
	\bibitem{BMU03}
		S. Borst, M. Mandjes, M. van Uitert (2003). Generalized processor sharing with light-tailed and heavy-tailed input. IEEE/ACM Transactions on Networking, 11, 821-834.
		
	\bibitem{BZ07}
		O. Boxma, B. Zwart (2007). Tails in scheduling. Performance Evaluation Review, 34, 13-20.

	\bibitem{BSS09}
		L. Bui, R. Srikant, A. Stolyar (2009). Novel architectures and algorithms for delay reduction in back-pressure scheduling and routing. In: Proc. Infocom 2009.
				
	\bibitem{ESP05}
		A. Eryilmaz, R. Srikant, J. Perkins (2005). Stable scheduling policies for fading wireless channels. IEEE/ACM Transactions on Networking, 13, 411-424.
		
	\bibitem{G96}
	  R. Gallager (1996). Discrete stochastic processes. Kluwer Academic.
	  
	\bibitem{GMT07}
		A. Ganti, E. Modiano, J. Tsitsiklis (2007). Optimal transmission scheduling in symmetric communication models with intermittent connectivity. IEEE Transactions on Information Theory, 53, 998-1008.
		
	\bibitem{GNT06}  
	  L. Georgiadis, M. Neely, L. Tassiulas (2006). Resource allocation and cross-layer control in wireless nertworks. Foundations and Trends in Networking, 1, 1-144.
		
	\bibitem{GW86}
		P. Glynn, W. Whitt (1986). A central-limit-theorem version of $L = \lambda W$. Queueing Systems, 1, 191-215.
	
	\bibitem{H06}
		B. Hajek (2006). Notes on communication network analysis. Available online at: http://www.ifp.illinois.edu/$\sim$hajek/Papers/networkanalysis Dec06.pdf.

	\bibitem{KS90}
	  P. R. Kumar, T. Seidman (1990). Dynamic instabilities and stabilization methods in distributed real-time scheduling of manufacturing systems. IEEE Transactions on Automatic Control, 35, 289-298.	
	  
	\bibitem{LTWW94}
	  W. Leland, M. Taqqu, W. Willinger, D. Wilson (1994). On the self-similar nature of ethernet traffic. IEEE/ACM Transactions on Networking, 2, 1-15.

	\bibitem{MMW89}
		A. Makowski, B. Melamed, W. Whitt (1989). On averages seen by arrivals in discrete time. In: Proc. CDC 1989.
			  
	\bibitem{MMT09}
	  M. Markakis, E. Modiano, J. Tsitsiklis (2009). Scheduling policies for single-hop networks with heavy-tailed traffic. In: Proc. Allerton 2009.
	  	  
	\bibitem{N08}
		M. Neely (2008). Order optimal delay for opportunistic scheduling in multi-user wireless uplinks and downlinks. IEEE/ACM Transactions on Networking, 16, 1188-1199.
  
	\bibitem{PW00}
	  K. Park, W. Willinger (2000). Self-similar network traffic: an overview. In: Self-Similar Network Traffic and Performance Evaluation, K. Park and W. Willinger, editors, Wiley Inc.

	\bibitem{RS92}
		A. Rybko, A. Stolyar (1992). Ergodicity of stochastic processes describing the operation of open queueing networks. Probl. Peredachi Inf., 3, 3-26.
		
	\bibitem{STZ11}
	  D. Shah, J. Tsitsiklis, Y. Zhong (2011). Optimal scaling of average queue sizes in an input-queued switch: an open problem. To appear in Queueing Systems.
	  			  
	\bibitem{SW06}
	  D. Shah, D. Wischik (2006). Optimal scheduling algorithms for input-queued switches. In: Proc. Infocom 2006.

	\bibitem{SW08}
	  D. Shah, D. Wischik (2008). Lower bound and optimality in switched networks. In: Proc. Allerton 2008.
	  
	\bibitem{SW93}
		K. Sigman, R. Wolff (1993). A review of regenerative processes. SIAM Review, 35, 269-288.
			  
	\bibitem{S04}
		A. Stolyar (2004). Maxweight scheduling in a generalized switch: state space collapse and workload minimization in heavy traffic. The Annals of Applied Probability, 14, 1-53.
	  
	\bibitem{TE92}
	  L. Tassiulas, A. Ephremides (1992). Stability properties of constrained queueing systems and scheduling policies for maximum throughput in multihop radio networks. IEEE Transactions on Automatic Control, 37, 1936-1948.
	  
	\bibitem{TE93}
		L. Tassiulas, A. Ephremides (1993). Dynamic server allocation to parallel queues with randomly varying connectivity. IEEE Transactions on Information Theory. 39, 466-478.
	
	\bibitem{WI91}
		D. Williams (1991). Probability with Martingales. Cambridge University Press.
	  
\end{thebibliography}
\end{document}